\documentclass[aps,prc,preprint,showkeys,superscriptaddress,amsmath,amssymb,floatfix]{revtex4}

\usepackage[dvips]{graphicx,color}
\usepackage{dcolumn}
\usepackage{bm}

\def\pmb#1{\setbox0=\hbox{#1}%
  \kern-.025em\copy0\kern-\wd0 
  \kern.05em\copy0\kern-\wd0
  \kern-.025em\raise.0433em\box0 }

\newcommand{\Figuretable}[1]{%
  \begin{center} --------- {\bf #1} --------- \\ \end{center}} 
\def\lambdabar{\protect\@lambdabar}
\def\@lambdabar{%
\relax
\bgroup
\def\@tempa{\hbox{\raise.73\ht0
\hbox to0pt{\kern.25\wd0\vrule width.5\wd0
height.1pt depth.1pt\hss}\box0}}%
\mathchoice{\setbox0\hbox{$\displaystyle\lambda$}\@tempa}%
{\setbox0\hbox{$\textstyle\lambda$}\@tempa}%
{\setbox0\hbox{$\scriptstyle\lambda$}\@tempa}%
{\setbox0\hbox{$\scriptscriptstyle\lambda$}\@tempa}%
\egroup
}

\begin{document}

\preprint{J-PARC-TH-0085}

\title{\boldmath
Repulsion and absorption of the $\Sigma$-nucleus potential for $\Sigma^-$-$^5$He
in the $^6$Li($\pi^-$,~$K^+$) reaction
}

\author{Toru~Harada}%
\email{harada@osakac.ac.jp}
\affiliation{%
Research Center for Physics and Mathematics,
Osaka Electro-Communication University, Neyagawa, Osaka, 572-8530, Japan
}
\affiliation{%
J-PARC Branch, KEK Theory Center, Institute of Particle and Nuclear Studies,
High Energy Accelerator Research Organization (KEK),
203-1, Shirakata, Tokai, Ibaraki, 319-1106, Japan
}

\author{Ryotaro~Honda}%
\affiliation{%
Department of Physics, Tohoku University, Sendai, Miyagi, 980-8578, Japan
}

\author{Yoshiharu~Hirabayashi}%
\affiliation{%
Information Initiative Center, 
Hokkaido University, Sapporo, 060-0811, Japan
}

\date{\today}

\begin{abstract}
We study phenomenologically inclusive spectra of 
the $^6$Li($\pi^-$,~$K^+$) reaction at 1.2 GeV/$c$ 
within a distorted-wave impulse approximation
with the optimal Fermi-averaging $\pi^-  p \to K^+ \Sigma^-$ $t$ matrix. 
We attempt to clarify the property of a $\Sigma$-nucleus potential for $\Sigma^-$-$^5$He 
by comparing the calculated spectra with the data of the J-PARC E10 experiment. 
The result shows that the repulsive and absorptive components 
of the $\Sigma^-$-$^5$He potential provide the ability to explain 
the data of the continuum spectra in $\Sigma$ and $\Lambda$ regions; 
the strengths of $V_\Sigma \simeq$ +30 MeV and $W_\Sigma \simeq$ $-$26 MeV 
are favored within the Woods-Saxon potential, consistent with analyses for heavier nuclei. 
Effects of the size and potential range for $\Sigma^-$-$^5$He in the neutron excess of
$(N-Z)/(N+Z)=$ 0.2 are also discussed. 
\end{abstract}
\maketitle


\section{Introduction}

Recently, the J-PARC E10 collaboration \cite{Sugimura14,Honda16} performed 
experimental measurements of the double charge-exchange (DCX) reaction
($\pi^-$,~$K^+$) on a $^6$Li target at $p_{\pi^-}=$ 1.2 GeV/$c$; 
missing mass spectra from $\Lambda$ to $\Sigma$ regions were obtained 
with $K^+$ forward-direction angles of $\theta_{\rm lab}=$ 2$^\circ$--14$^\circ$, 
whereas no significant peak structure of a neutron-rich $^6_\Lambda$H hypernucleus 
was observed around the $^4_\Lambda{\rm H}+2n$ threshold \cite{Sugimura14,Honda16}.
This reaction can also populate $\Sigma^-\otimes{^5{\rm He}}$ doorway states 
with $T=$ 3/2 in a $^6_\Sigma$H hypernucleus by a $\pi^- p \to K^+ \Sigma^-$ 
process in the nuclear medium \cite{Harada17}. 
One expects that a $\Sigma$-nucleus potential for $\Sigma^-$-$^5$He 
can be studied by comparing a theoretical calculation with the data of 
the missing mass spectra at $\theta_{\rm lab}=$ 2$^\circ$--14$^\circ$ 
in the reaction \cite{Honda16}. 

The DCX ($\pi^-$,~$K^+$) reactions on nuclear targets provide 
investigation for the $\Sigma$-nucleus 
potential analyzing quasi-free (QF) $\Sigma^-$ production spectra. 
Noumi and his collaborators \cite{Noumi02,Saha04} performed 
measurements of $\Sigma$-hypernuclei by inclusive ($\pi^-$,~$K^+$) 
reactions on heavier targets at $p_{\pi^-}=$ 1.2 GeV/$c$ in the KEK-E438 experiment.
Their analysis within a distorted-wave impulse approximation (DWIA) 
indicated that the $\Sigma$-nucleus potential has a strong repulsion 
in the real part and a sizable absorption in the imaginary part \cite{Saha04}.
Batty and his collaborators \cite{Batty94,Friedman07} studied the $\Sigma$-nucleus potential 
analyzing $\Sigma^-$ atomic x-ray data systematically.
Thus the latest studies \cite{Batty94,Friedman07,Mares95,Dabrowski02,Harada05,Harada06} 
have suggested that the $\Sigma$-nucleus potential has a repulsion inside 
the nuclear surface and a shallow attraction outside the nucleus 
with a sizable absorption, e.g., 
the density-dependent (DD) potential \cite{Batty94}.
This repulsion may originate from $\Sigma N$ $I=$ 3/2, $^{3}S_1$ 
channel \cite{Dover89,Harada90,Harada98}, 
whose state corresponds to a quark Pauli-forbidden state 
in a baryon-baryon system \cite{Oka83}, 
as supported by modern $YN$ potentials \cite{Rijken10} and 
also recent Lattice QCD calculations \cite{LQCD11}.

Harada and Hirabayashi \cite{Harada05,Harada06} 
succeeded in explaining the data of the ($\pi^-$,~$K^+$) reaction at 1.2 GeV/$c$ 
on $^{28}$Si and $^{209}$Bi targets, performing DWIA calculations with the optimal
Fermi averaging for the $\pi^-  p \to K^+  \Sigma^-$ reaction.
Their analysis with the DWIA \cite{Harada05,Harada06} also suggested that 
the $\Sigma$-nucleus potential 
within the Woods-Saxon (WS) or 2pF form is 
\begin{equation}
U_\Sigma(r)= (V_\Sigma + i W_\Sigma)/[1 + \exp{((r-R)/a)}]
\label{eqn:e2}
\end{equation}
with $R=1.1A_{\rm core}^{1/3}$ and $a=$ 0.67 fm, 
where $V_\Sigma=$ ($+20$)--($+30$) MeV and $W_\Sigma=$ ($-20$)--($-40$) MeV,
corresponding to a strong repulsion in the real part and a sizable absorption 
in the imaginary part of the potential. 
It is very important to study a negatively charged $\Sigma^-$ hyperon 
in the nuclear medium in order to obtain valuable information concerning the maximal 
mass of neutron stars, 
in which a baryon fraction is found to depend on properties of the $\Sigma^-$ 
potentials for astrophysics \cite{Balberg97}.
The $\Sigma$-nucleus interaction may be established as being repulsive so far, 
but it is still an open problem how repulsive 
the $\Sigma$-nucleus potential is \cite{Gal16,Inoue16}.

In this paper, we investigate phenomenologically the inclusive spectra of 
$\Sigma$ hypernuclear production by the $^6$Li($\pi^-$,~$K^+$) reaction 
at 1.2 GeV/$c$ 
in order to extract valuable information on the $\Sigma$-nucleus (optical) 
potential for $\Sigma^-$-$^5$He from the data of the J-PARC E10 experiment \cite{Honda16}.
We demonstrate the calculated spectra within the DWIA, 
using the optimal Fermi-averaging $t$ matrix for 
the $\pi^-  p \to K^+  \Sigma^-$ reaction in the nuclear medium 
because the energy dependence of the $t$ matrix is important to explain 
the behavior of the ($\pi^-$, $K^+$) spectrum \cite{Harada05}.
By using a single-particle $\Sigma^-\otimes{^5{\rm He}}$ model 
with a spreading potential, 
we study the nature of the repulsion and absorption in the $\Sigma$-nucleus 
potential, in comparison with the data of the J-PARC E10 experiment.

\section{Calculations}

\subsection{Distorted wave impulse approximation}

\begin{figure}[htb]
  \begin{center}
  \includegraphics[width=0.80\linewidth]{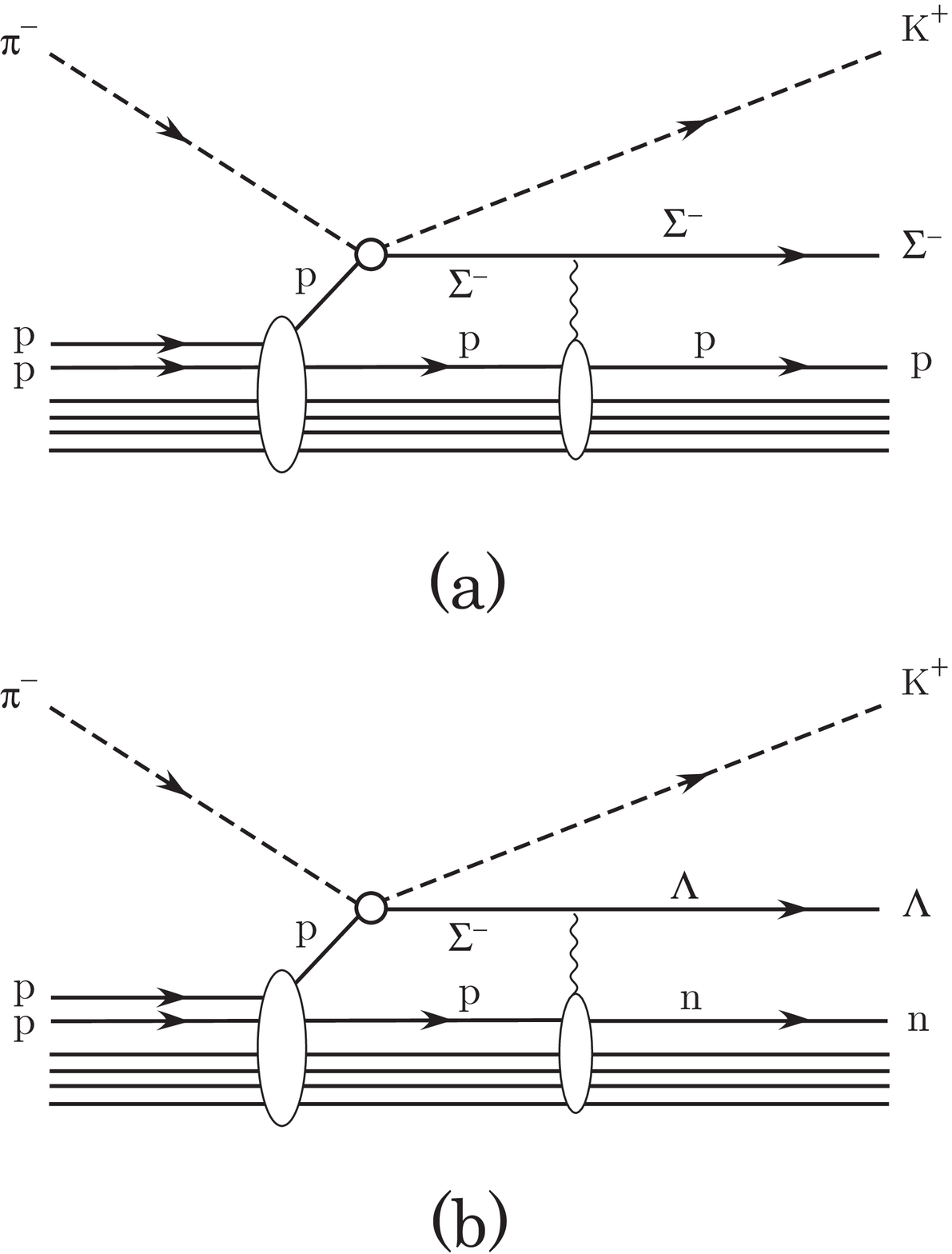}
  \caption{\label{fig:1}
  Diagrams for double-charge exchange ($\pi^-$,~$K^+$) reactions on nuclear targets, 
  leading to production of (a) $\Sigma$ hypernuclear states by $\pi^-p$ $\to$ $K^+ \Sigma^-$ processes  
  and (b) $\Lambda$ hypernuclear states caused by the 
  $\Sigma^-p$ $\leftrightarrow$ $\Lambda n$ coupling.
  }
  \end{center}
\end{figure}

Let us consider production of $\Sigma$ hypernuclear states in 
the DCX reaction ($\pi^-$,~$K^+$) on a nuclear target. 
Figure~\ref{fig:1} illustrates diagrams for the nuclear ($\pi^-$,~$K^+$) reaction 
by $\pi^-p$ $\to$ $K^+ \Sigma^-$ processes for the $\Sigma$ hypernuclear states, 
and by $\pi^-p$ $\to$ $K^+ \Sigma^-$ via $\Sigma^-$ doorways caused by 
the $\Sigma^-p \to \Lambda n$ conversion for $\Lambda$ hypernuclear excited states.
According to the Green's function method \cite{Morimatsu94} in the DWIA, 
an inclusive $K^+$ double-differential laboratory cross section 
of the $\Sigma$ production on the nuclear target with a spin 
$J_A$ (its $z$-component $M_A$) \cite{Hufner74}
is given by
\begin{equation}
{{d^2\sigma} \over {d\Omega dE} } 
 = {1 \over {[J_A]}} \sum_{M_A}S(E) 
\label{eqn:e3}
\end{equation}
with $[J_A]=2J_A+1$. The strength function $S(E)$ is 
written as
\begin{eqnarray}
S(E)&=&-{1 \over \pi}{\rm Im} \sum_{\alpha \alpha'}
\int d{\bm r}d{\bm r'}
F_{\Sigma}^{\alpha \, \dagger}({\bm r}) 
{G}_{\Sigma}^{\alpha\alpha'}(E;{\bm r},{\bm r}')\nonumber\\
&& \times 
F_{\Sigma}^{\alpha'}({\bm r}'),
\label{eqn:e4}
\end{eqnarray}
where 
${G}_{\Sigma}^{\alpha\alpha'}$ is a complete Green's function, 
$F_{\Sigma}^{\alpha}$ is a $\Sigma$ production amplitude
defined by
\begin{equation}
  F_{\Sigma}^{\alpha} = 
  \beta^{1 \over 2}{\overline{f}}_{\pi^-p \to K^+\Sigma^-}
  \chi_{{\bm p}_{K}}^{(-) \ast}
  \chi_{{\bm p}_{\pi}}^{(+)} 
  \langle \alpha \, | \hat{\mit\psi}_p | \, \mit\Psi_A \rangle,
\label{eqn:e5}
\end{equation}
and $\langle \alpha \, | \hat{\mit\psi}_p  | \, \mit\Psi_A \rangle$ 
is a hole-state wave function for a struck proton in the target; 
$\alpha$ ($\alpha'$) denotes the complete set of eigenstates for the system.
The laboratory energy and momentum transfer is
\begin{equation}
   \omega= E_{K}-E_{\pi},  \qquad  {\bm q}={\bm p}_K - {\bm p}_{\pi},
\label{eqn:e6}
\end{equation}
where $E_{K}$ and ${\bm p}_K$ ($E_{\pi}$ and ${\bm p}_{\pi}$) denote 
the energy and momentum of the outgoing $K^+$ (the incoming $\pi^-$),
respectively. 
The kinematical factor $\beta$ denotes the translation 
from the two-body $\pi^-$-$p$ laboratory system to the $\pi^-$-$^{6}$Li 
laboratory system. 
$\overline{f}_{\pi^-p \to K^+\Sigma^-}$ is a  
Fermi-averaged amplitude for the $\pi^-p \to K^+\Sigma^-$ reaction 
in the nuclear medium \cite{Harada04,Harada05,Harada06}.
$\chi_{{\bm p}_{K}}^{(-)}$ and $\chi_{{\bm p}_{\pi}}^{(+)}$ 
are distorted waves for the outgoing $K^+$ and 
incoming $\pi^-$ mesons, respectively.

\Figuretable{FIG. 1}

\subsection{$^6$Li target}
\label{6Li}

For the $^{6}$Li target, 
we assume single-particle (s.p.) description of a proton 
for simplicity \cite{Dover82}, 
although the state of $^{6}$Li($1^+_{\rm g.s.}$; $T$=0) 
is well described as $\alpha +d$ clusters \cite{Fujiwara80}.
Thus a s.p.~wave function for the proton in 
$0p_{3/2}$ ($0s_{1/2}$) is calculated by the WS potential with 
$a=$ 0.67 fm, $R=$ $1.27A^{1/3}$ = 2.31 fm  \cite{Bohr69}.
The strength parameter of the potential must be adjusted to 
be $V^N_0=$ $-$55.5 MeV ($-$58.0 MeV) for the proton in the $p_{3/2}$ 
($s_{1/2}$) state, and $V^N_{\rm so}=$ $-0.44 V_0^N$, 
in order to reproduce the data of proton s.p.~energies 
in $^{6}$Li($p$,~2$p$) reactions \cite{Tilley02,Jacob66}. 
Thus we obtain the s.p.~energies of $-$4.61 MeV for $0p_{3/2}$
and $-$21.48 MeV for $0s_{1/2}$, and the charge radius 
for $^{6}$Li($1^+_{\rm g.s.}$) becomes 2.48 fm.
This value in the s.p.~model is slightly smaller than 
that of $2.56 \pm 0.05$ fm in electron elastic scatterings \cite{Vries87} 
because the structure of $\alpha + d$ clusters is not taken 
into account \cite{Fujiwara80}.

\subsection{Eikonal distortion}

Due to a high momentum transfer 
$q \simeq$ 320--600 MeV/$c$ in the ($\pi^-$,~$K^+$) reaction
at $K^+$ forward-direction angles of 
$\theta_{\rm lab}=$ 2$^\circ$--14$^\circ$, 
we simplify the computational procedure for the distorted waves, 
$\chi_{{\bm p}_{K}}^{(-)}$ and $\chi_{{\bm p}_{\pi}}^{(+)}$, 
using the eikonal approximation. 
To reduce ambiguities in the distorted-waves, 
we adopt the same parameters used in calculations for the
$\Lambda$ and $\Sigma^-$ QF spectra in nuclear ($\pi^\mp$,~$K^+$) 
reactions \cite{Harada05,Harada06,Harada04}.
Here we used total cross sections of $\sigma_\pi$= 32 mb 
for $\pi^- N$ scattering and $\sigma_K$= 12 mb for $K^+ N$ one, 
and $\alpha_\pi = \alpha_K =$ 0, as the distortion parameters.
We took into account the recoil effects 
because the effects are very important to estimate the production 
spectra for the light nuclear system, 
leading to an effective momentum transfer having 
$q_{\rm eff} \simeq (1-1/A)q \simeq 0.83 q$ for $A=$ 6.

\subsection{Green's functions}

To calculate the nuclear ($\pi^-$,~$K^+$) spectra in the DWIA, 
we employ the Green's function method \cite{Morimatsu94}, 
which is one of the most powerful treatments in the calculation 
of a spectrum in which not only bound states but also 
continuum states are described with an absorptive potential 
for spreading components.
The complete Green's function $G(E)$ has all information 
concerning $\Sigma^- \otimes {^5}{\rm He}$ dynamics, and it 
can be obtained by solving the following equation numerically:
\begin{equation}
{G_\Sigma}(E)=G_\Sigma^{(0)}(E)
+G_\Sigma^{(0)}(E)U_\Sigma(E){G_\Sigma}(E),
\label{eqn:e7}
\end{equation}
where $G^{(0)}$ is the free Green's function. 
The $\Sigma$-nucleus (optical) potential is given by
\begin{equation}
 U_\Sigma(E)=PUP+ PUQ G(E+i\epsilon) QUP, 
\label{eqn:e8}
\end{equation}
where $U$ is hyperon-nucleus interaction and $P+Q=1$ in the Feshbach 
projection method. 
The strength function $S(E)$ in Eq.~(\ref{eqn:e4})
can be evaluated by taking the complete Green's function 
${G}_{\Sigma}$ in Eq.~(\ref{eqn:e7}) which fully includes 
$\Sigma^-$ doorways by $\pi^- p \to K^+ \Sigma^-$ reactions \cite{Harada09},
as shown in Fig.~\ref{fig:1}.
Because non-spin-flip processes seem to dominate in 
$\pi^- p \to K^+ \Sigma^-$ reactions at 1.2 GeV/$c$ \cite{Kamano13}, 
configurations of $\Sigma^- \otimes {^5{\rm He}}$ with $T=$3/2, 
$J^\pi=$ $(1^+ \otimes \Delta L)=$ 
1$^+$, 0$^-$, 1$^-$, 2$^-$, 1$^+$, 2$^+$, 3$^+$, $\cdots$, 
can be populated where $\Delta L=$ 0, 1, 2, $\cdots$, denote 
the angular momentum transfer to $^6{\rm Li}$(1$^+_{\rm g.s.}$).
No $\Lambda$ channel is explicitly taken into account 
because the $\Sigma\Lambda$ coupling effects may be described 
as a spreading imaginary potential in continuum spectra in 
$\Sigma$ regions.

\section{Optimal Fermi averaging}

As discussed in Refs.~\cite{Harada05,Harada06}, the DWIA analyses of 
the $\Sigma^-$ QF spectra in the nuclear ($\pi^-$,~$K^+$) 
reactions indicated importance of an energy dependence 
of the Fermi-averaged amplitude
${\overline{f}}_{\pi^-p \to K^+\Sigma^-}$ 
in order to extract properties of the potential from the data.
In this version of the DWIA, 
the Fermi-averaged amplitude $\overline{f}_{\pi^-p \to K^+\Sigma^-}$ 
in Eq.~(\ref{eqn:e5}) plays 
an important role in explaining a spectral shape  
in the nuclear ($\pi^-$,~$K^+$) reaction \cite{Harada05,Harada06}. 
We should use the $\pi^-p \to K^+\Sigma^-$ $t$ matrix 
which can fully reproduce the experimental data of differential cross 
sections in free space so as to obtain a suitable Fermi-averaged amplitude 
in our calculations.
Thus we perform the optimal Fermi averaging 
for elementary $\pi^-  p \to K^+  \Sigma^-$ processes at each 
$\theta_{\rm lab}$ in the nucleus \cite{Harada05,Harada06}.

\subsection{${\bm \pi}^- {\bm p}\to {\bm K}^+ {\bm \Sigma}^-$ reactions}

Very recently, new data of 4 angular points 
for the $\pi^- p \to K^+ \Sigma^-$ reactions
have been measured with excellent quality at $E_{\rm c.m.}=$ 1875 MeV 
in the J-PARC E10 experiment \cite{Honda16}.
Hence we improve the angular distributions of the differential 
cross section in the center-of-mass (c.m.) frame, 
whereas there is still no available amplitude of the $\pi^- p \to K^+ \Sigma^-$ 
reaction due to poor quality in the other data \cite{Dahl67}.
The cross section is written as 
\begin{eqnarray}
\Bigl( {d \sigma \over d \Omega} \Bigr)^{\rm elem}_{\rm cm}
&=&
\lambdabar^2 \sum_{\ell=0}^{\ell_{\rm max}} A_{\ell}(E_{\rm cm})
 P_\ell(\cos{\theta_{\rm cm}})    \nonumber\\
& =&
{\omega_f  \omega_i p_f \over (2\pi)^2 p_i}
|t_{cm}(E_{\rm cm})|^2,
\label{eqn:e9}
\end{eqnarray}
where $\lambdabar$ is the de Broglie wavelength of $\pi^- p$, 
and $P_\ell(x)$ are Legendre polynomials. 
Coefficient parameters  $A_{\ell}(E_{\rm cm})$ are expressed 
by a power series of $E_{\rm cm}$ so as to make a fit to 
their energy dependence.
$t_{cm}(E_{\rm cm})$ denotes the elementary $\pi^- p \to K^+ \Sigma^-$
$t$ matrix in the c.m.~frame, and 
$p_f$ ($p_i$) and $\omega_f$ ($\omega_i$) are a momentum and a reduced energy 
for $K^+ \Sigma^-$ ($\pi^- p$) in the c.m.~frame,  respectively. 

\begin{figure*}[thb]
\begin{center}
  \includegraphics[width=1.0\linewidth]{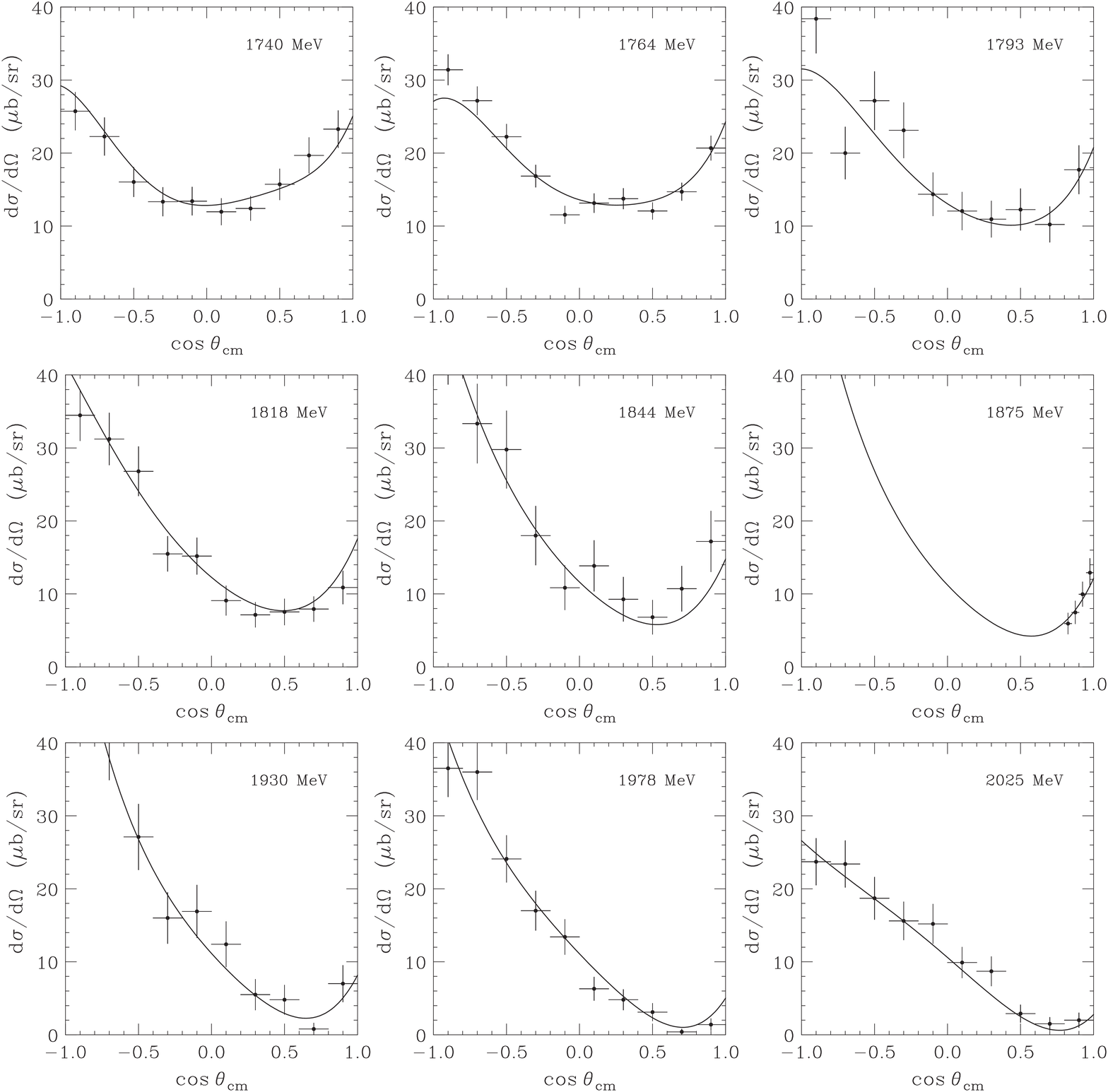}
\end{center}
\caption{\label{fig:2}
  Calculated angular distributions of the differential cross section for 
  the $\pi^- p \to K^+ \Sigma^-$ reaction in the c.m.~frame at 
  $E_{\rm cm}$ = 1740, 1764, 1793, 1818, 1844, 1875, 1930, 1978, and 2025 MeV, 
  together with the data \cite{Dahl67,Honda16}. 
}
\end{figure*}

In Fig.~\ref{fig:2}, we show the angular distributions 
at various c.m. energies $E_{\rm cm}$, 
together with the data \cite{Dahl67,Honda16}.
The angular distributions near $E_{\rm cm}=$ 1875 MeV
are improved for fits to the data newly observed 
in the J-PARC E10 experiment \cite{Honda16}, 
whereas the value of an integrated cross section 
$\sigma_{\rm cm}^{\rm tot}(E_{\rm cm})$ is not so changed 
from that of the previous one \cite{Harada05,Harada06}.

\Figuretable{FIG. 2}

\subsection{Optimal Fermi-averaged $t$ matrix}

To fully describe the energy dependence of the ($\pi^-$,$K^+$) reaction
on a target nucleus, which comes from $N^*$ resonances in 
the $\pi^- p \to K^+ \Sigma^-$ processes, 
we must adopt the optimal Fermi averaging \cite{Harada04} for the $\pi^- p \to K^+ \Sigma^-$
reaction in the nucleus.
In this version of the DWIA, 
an ``optimal'' cross section for the elementary $\pi^- p \to K^+ \Sigma^-$ processes
in the nucleus \cite{Harada05} can be given as 
\begin{eqnarray}
\Bigl( \displaystyle{d \sigma \over d \Omega} \Bigr)^{\rm opt}
&\equiv& 
\beta |\overline{f}_{\pi^-p \to K^+\Sigma^-}|^2 \nonumber\\
& =&
{ p_K E_K \over (2\pi)^2 v_\pi}|t^{\rm opt}_{\pi N,K \Sigma}(p_\pi; \omega,{\bm q})|^2,
\label{eqn:e10}
\end{eqnarray}
where $v_\pi=p_\pi/E_\pi$, and 
${t}^{\rm opt}_{\pi N,K \Sigma}(p_\pi; \omega,{\bm q})$ is 
an optimal Fermi-averaged $\pi^-p \to K^+\Sigma^-$ $t$ matrix, 
which is defined by 
\begin{widetext}
\begin{equation}
{t}^{\rm opt}_{\pi N,K \Sigma}(p_\pi; \omega,{\bm q}) =
{\int^{\pi}_{0} \sin{\theta_N} d \theta_N \int_{0}^{\infty} 
dp_{N} p_N^2 \rho{(p_N)}{t}_{\pi N,K \Sigma}(E_{2};{\bm p}_\pi,{\bm p}_N)
\over
\int^{\pi}_{0} \sin{\theta_N} d \theta_N \int_{0}^{\infty} 
dp_{N} p_N^2 \rho{(p_N)}
  }\Biggl|_{{\bm p}_N={\bm p}^*_N}, 
\label{eqn:e11}
\end{equation}
\end{widetext}
where $E_N$ and ${\bm p}_N$ are the energy and momentum of a proton
in the target nucleus, respectively; 
$\cos{\theta_N}= \hat{\bm p}_\pi\cdot\hat{\bm p}_N$, 
$E_{2}=E_{\pi}+E_{N}$ is a total energy of the $\pi N$ system, 
and $\rho(p_N)$ is a Fermi-momentum distribution of the proton in the target nucleus. 
The momentum ${\bm p}_N^*$ in Eq.~(\ref{eqn:e11}) is a solution which satisfies 
the on-energy-shell equation for a struck proton in the nuclear systems, 
\begin{eqnarray}
\sqrt{({\bm p}_N^*+{\bm q})^2+m_\Sigma^2}-\sqrt{({\bm p}_N^*)^2+m_N^2}=\omega,
\label{eqn:e12}
\end{eqnarray}
where $m_\Sigma$ and $m_N$ are masses of the $\Sigma^-$ and the 
proton, respectively. 
This procedure constructed from the on-energy-shell 
$\pi^- p \to K^+ \Sigma^-$ processes in the nucleus \cite{Gurvitz86} guarantees
to have optimal values for $\overline{f}_{\pi^-p \to K^+\Sigma^-}$
in a factorized form of Eq.~(\ref{eqn:e5}).
Here we neglected the energy-dependence of a phase in the
$\pi^-p \to K^+\Sigma^-$ $t$ matrix, and replaced 
${t}_{\pi N,K \Sigma}(E_{2};{\bm p}_\pi,{\bm p}_N)$ in the laboratory frame
by its absolute value 
$|{t}_{\pi N,K \Sigma}(E_{2};{\bm p}_\pi,{\bm p}_N)|$  which is obtained 
from the corresponding one in the c.m.~frame in Eq.~(\ref{eqn:e9}); 
${t}_{\pi N,K \Sigma}(E_{2};{\bm p}_\pi,{\bm p}_N)= \eta\, t_{cm}(E_{\rm cm})$,
where $\eta$ is the M\"oller factor.
Such an assumption has been confirmed to be appropriate in the case of
the $\pi^+ n \to K^+ \Lambda$ reaction \cite{Harada04}, 
leading to the $\omega$ dependence of $({d \sigma/d \Omega})^{\rm opt}$ which 
is significant to describe the behavior of the ($\pi^\pm$,$K^+$) reactions \cite{Harada05}.

\begin{figure}[tbh]
  \begin{center}
  \includegraphics[width=1.00\linewidth]{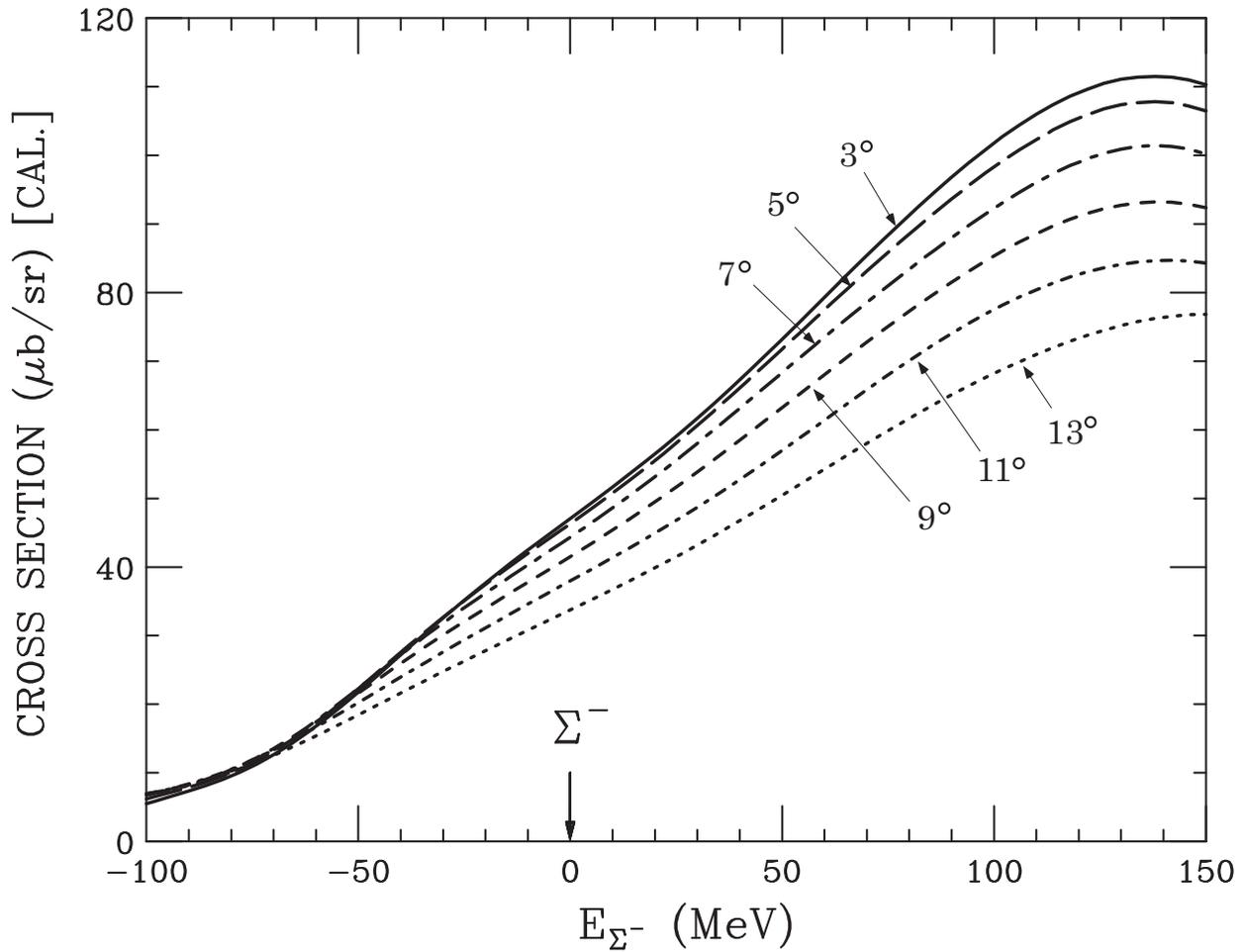}
  \end{center}
  \caption{\label{fig:3}
  Energy dependence of the optimal cross section $(d \sigma/d \Omega)^{\rm opt}$
  for the $\pi^- p \to K^+ \Sigma^-$ reaction on the $^6$Li target
  at $p_{\pi^-}$=1.2 GeV/$c$ and $\theta_{\rm lab}=$ 3$^\circ$, 5$^\circ$, 
  7$^\circ$, 9$^\circ$, 11$^\circ$, and 13$^\circ$, as a function of $E_{\Sigma^-}$. 
  The arrow shows the $^5{\rm He}$+$\Sigma^-$ threshold.
  }
\end{figure}

In Fig.~\ref{fig:3}, we show the optimal Fermi-averaged cross sections
of $({d \sigma/d \Omega})^{\rm opt}$ 
at $\theta_{\rm lab}=$ 3$^\circ$, 5$^\circ$, 7$^\circ$, 
9$^\circ$, 11$^\circ$, and 13$^\circ$ in the region from $\Lambda$ to $\Sigma^-$. 
We confirm that there appears a strong energy dependence of 
the $\pi^-p \to K^+\Sigma^-$ reaction in the nuclear medium,  
together with the angular dependence of $\theta_{\rm lab}$. 
Therefore, such behavior of $(d \sigma/d \Omega)^{\rm opt}$ 
would play a significant role in explaining the shape 
of the spectrum in the nuclear ($\pi^-$,~$K^+$) reaction, 
as discussed in Refs.~\cite{Harada05,Harada06}.
Because $\overline{f}_{\pi^-p \to K^+\Sigma^-}$ directly 
affects the spectral shape including widely the $\Sigma^-$ QF region, 
thus one should carefully extract information concerning 
the $\Sigma$-nucleus potential from the data.

\Figuretable{FIG. 3}

\section{${\bm \Sigma}$-nucleus potential}
\label{Sigmapot}

The $\Sigma$-nuclear final states are obtained by solving 
the Schr\"odinger equation
\begin{eqnarray}
&& \left[-{\hbar^2 \over 2\mu}\nabla^2 
+ U_\Sigma(r)+ U_{\rm Coul}(r)  \right] \Psi_\Sigma 
= E \Psi_\Sigma,
\label{eqn:e1}
\end{eqnarray}
where $\mu$ is the $\Sigma$-nucleus reduced mass, 
$U_\Sigma$ is the $\Sigma$-nucleus potential, and $U_{\rm Coul}$ 
is the Coulomb potential.
Here the $\Sigma$-nucleus potential for $\Sigma^-$-$^5{\rm He}$
is given as
\begin{equation}
U_{\Sigma}(r)= [V_{\Sigma}+iW_{\Sigma}g(E_\Lambda)]f(r)
\label{eqn:e13}
\end{equation} 
with the assumption of the WS form
\begin{eqnarray}
f(r)=[1 + \exp{((r-R)/a)}]^{-1},
\label{eqn:e14}
\end{eqnarray} 
where $R= r_0A^{1/3}_{\rm core}$ and $a$ denote the radius 
and diffuseness of the potential, respectively, in order to be compared with 
the $\Sigma$-nucleus potentials for $\Sigma^-$-$^{27}$Al \cite{Harada05} 
and $\Sigma^-$-$^{208}$Pb \cite{Harada06};
$g(E_\Lambda)$ is an energy-dependent function 
which linearly increases
from 0.0 at $E_\Lambda=$ 0 MeV to 1.0 at $E_\Lambda=$ 60 MeV
with respect to the $\Lambda$ emitted threshold, 
as often used in nuclear optical models.
The ground state of $^5$He as the nuclear core is known to be a 3/2$^-$ resonant state 
with the width of $\mit\Gamma=$ 0.65 MeV at the energy of $E_r=$ 0.80 MeV with 
respect to the $\alpha + n$ threshold \cite{Tilley02}.
Thus the appropriate parameters of ($r_0$,~$a$) in Eq.~(\ref{eqn:e14})
must be used.
For $U_{\rm Coul}$, we use an attractive Coulomb potential 
with the uniform distribution of a charged sphere where
$ Z = 2$ for $\Sigma^-$-$^5$He.

\subsection{Folding-model potential}
\label{Folding}

\begin{figure}[tbh]
  \begin{center}
  \includegraphics[width=1.00\linewidth]{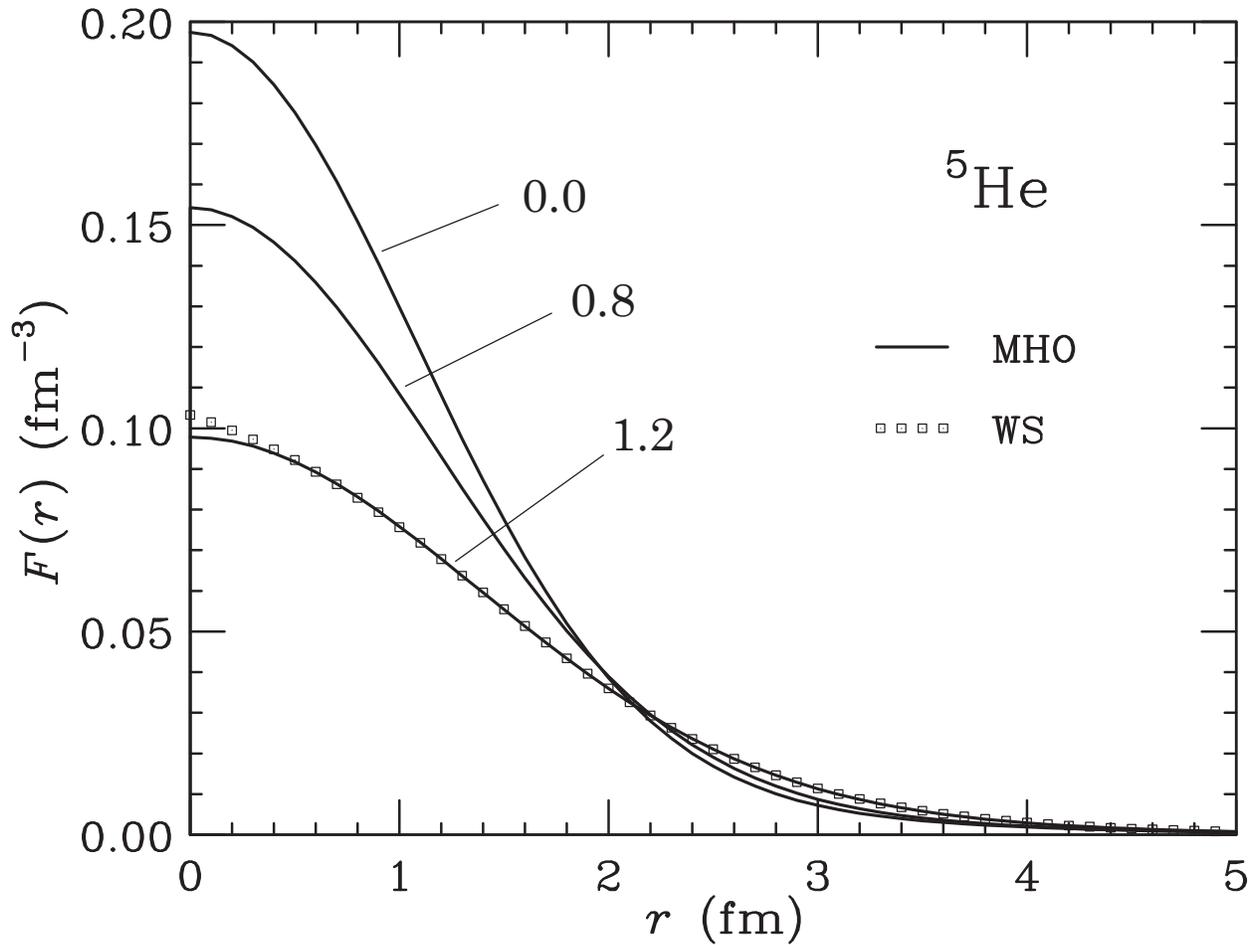}
  \end{center}
  \caption{\label{fig:4}
  Radial distributions of the form factors with Gaussian ranges of 0.0, 0.80, and 1.20 fm 
  for the $^5$He(3/2$_{\rm g.s.}^-$) nucleus, as a function of the radial distance. 
  Solid curves and square symbols denote the distributions with 
  the modified harmonic oscillator (MHO) and the Woods-Saxon (WS) models, respectively. 
  }
\end{figure}

To determine the parameters of ($r_0$,~$a$) for the nuclear core 
in the WS form, we consider a folding-model potential obtained by 
convoluting the nuclear one-body density with a two-body $\Sigma^- N$ 
force \cite{Greenlees68}.
The folding-model potential is given by  
\begin{eqnarray}
U_\Sigma(r)= \int v_{\Sigma N}({\bm r}-{\bm r'})\rho({\bm r'})d{\bm r'}, 
\label{eqn:e15}
\end{eqnarray}
where $\rho(r)$ denotes the nuclear density distribution 
normalized by 
\begin{eqnarray}
4 \pi \int_0^\infty \rho(r)r^2 dr=A_{\rm core}.
\label{eqn:e16}
\end{eqnarray}
Because $^5$He(3/2$^-_{\rm g.s.}$) is a $p$-wave resonant state with 
a narrow width ($\mit\Gamma =$ 0.65 MeV), the corresponding wave function 
may behave approximately as a bound-state one. 
Thus we assume the s.p.~density of the shell model within 
bound-state approximation for simplicity; 
the modified harmonic oscillator (MHO) model is often used in the systematic 
description of the size and density distribution for He isotopes 
with $A=$ 4, 6, and 8 \cite{Tanihata13}.
We choose carefully the MHO size parameters of $b_s=$ 1.71 fm and $b_p=$ 2.66 fm 
for $^5$He(3/2$^-_{\rm g.s.}$) with center-of-mass and nucleon-size 
corrections \cite{Uberall71}, providing the matter root-mean-square (rms) 
radius of $\langle r^2 \rangle^{1/2}=$ 2.43 fm which is obtained 
by the 3/2$^-$ resonant-state wave function calculated in the same procedure of Ref.~\cite{Sakuragi86}. 
For the two-body $\Sigma^- N$ force involving absorption, we assume a simple gaussian form,
\begin{equation}
v_{\Sigma N}(r)=(\bar{v}_{\Sigma N}+i\bar{w}_{\Sigma N})\exp{(-r^2/a_{\Sigma N}^2)},
\label{eqn:e17}
\end{equation}
where 
$\bar{v}_{\Sigma N}$ and $\bar{w}_{\Sigma N}$ are the real and imaginary parts 
of the $\Sigma N$ spin-isospin averaged strength, respectively; the range 
of $a_{\Sigma N}$ is chosen to be 1.2 fm, 
consistent with the range of a hyperon-nucleon potential in free space \cite{Batty94},
e.g., the D2$'$ potential \cite{Akaishi00}.
We define a nuclear form factor as 
\begin{eqnarray}
F(r)&=&\int \rho_G({\bm r}-{\bm r}')\rho({\bm r'})d{\bm r'}, 
\label{eqn:e18}
\end{eqnarray}
where $\rho_G(r)=(\sqrt{\pi}a_{\Sigma N})^{-3}\exp{(-r^2/a_{\Sigma N}^2)}$ 
which is normalized as $4 \pi \int \rho_G(r) r^2dr =1$.
Thus 
we have
\begin{eqnarray}
U_\Sigma(r)&=& (\bar{v}_{\Sigma N}+i\bar{w}_{\Sigma N})(\sqrt{\pi}a_{\Sigma N})^3F(r),  
\label{eqn:e19}
\end{eqnarray}
where $4 \pi \int F(r) r^2dr =A_{\rm core}$. 
Figure~\ref{fig:4} shows the form factors of $F(r)$ with 
$a_{\Sigma N}=$ 1.2, 0.8, and 0.0 fm.
For zero range ($a_{\Sigma N}=$ 0.0 fm), 
$F(r)$ is equal to the matter MHO density distribution of 
$^5$He(3/2$^-_{\rm g.s.}$). 
Note that $F(r)$ is reduced at the nuclear center 
because the radial distribution of the form factor 
is modified by $a_{\Sigma N}$ 
due to the small size of the nucleus.

\Figuretable{FIG. 4}

\subsection{Woods-Saxon parameters}
\label{WSpot}

For $\Sigma^-$-$^5{\rm He}$ in the present work, 
we use the WS form with the parameters of ($r_0$,~$a$) adjusted 
to give a best least-squares fit to the radial shape of the form factor 
obtained by folding a gaussian range of $a_{\Sigma N}=$ 1.2 fm 
into the matter MHO density distribution. 
The parameters of the resulting WS form in Eq.~(\ref{eqn:e14}) are 
\begin{equation}
r_0= 0.835 \, \mbox{fm}, \quad  a= 0.706 \, \mbox{fm},
\label{eqn:e19}
\end{equation}
which reproduce the radial shape of the form factor very well, 
as seen in Fig.~\ref{fig:4}; 
the rms radius of the potential \cite{Greenlees68,Sakaguchi13} denotes 
\begin{equation}
\langle r^2 \rangle^{1/2}_U=
 {\int r^2 U_\Sigma(r) d{\bm r}  \over  \int U_\Sigma(r) d{\bm r}}= 2.84 \ \ \mbox{fm}.
\label{eqn:e20}
\end{equation}
A spin-orbit potential for $\Sigma$ is also considered to denote a term of 
$V^{\Sigma}_{\rm so}
{(1/r)}{[df(r)/dr]}{\bm \sigma}{\cdot}{\bm L}$, 
where $V^{\Sigma}_{\rm so}\simeq$ ${1 \over 2}V^{N}_{\rm so}$
$\simeq$ 10 MeV \cite{Morimatsu84}.

The strength parameters of ($V_\Sigma$,~$W_\Sigma$) should 
be adjusted appropriately to reproduce available experimental data.
A spreading imaginary potential $W_\Sigma$ can represent 
complicated continuum states of ${^{6}_{\Lambda}}{\rm H}^*$. 
It should be noticed that the nuclear structure of $^5$H$(1/2^+_{\rm g.s.})$
as the core nucleus is rather uncertain \cite{Gal13,Hiyama13,Wuosmaa17}
although a resonant state at $E_{\rm ex}\simeq$ 1.7 MeV has been identified in 
Ref.~\cite{Korsheninnikov01}.
We assume that the $^5$H core state with ${\mit\Gamma}\simeq$ 2 MeV 
is located at $E_{\rm ex}=$ 4.0 MeV above the $^3{\rm H}+2n$ threshold, 
as suggested in Refs.~\cite{Tilley02,Harada17}.
Thus we have the $\Lambda$ emitted threshold ($E_\Lambda=$ 0 MeV)
corresponding to the $^4_\Lambda{\rm H}(1^+_{\rm exc.})+2n$ threshold
at $M_{\rm x}=$ 5802.79 MeV/$c^2$, so that the threshold-energy difference 
between $\Sigma^-$-$^{5}$He and $\Lambda$-$^{5}$H channels becomes  
$\Delta M = M({^5{\rm He}})+m_{\Sigma^-} -M({^5{\rm H}})-m_\Lambda=$ 
56.9 MeV.

\begin{figure}[tbh]
  \begin{center}
  \includegraphics[width=0.8\linewidth]{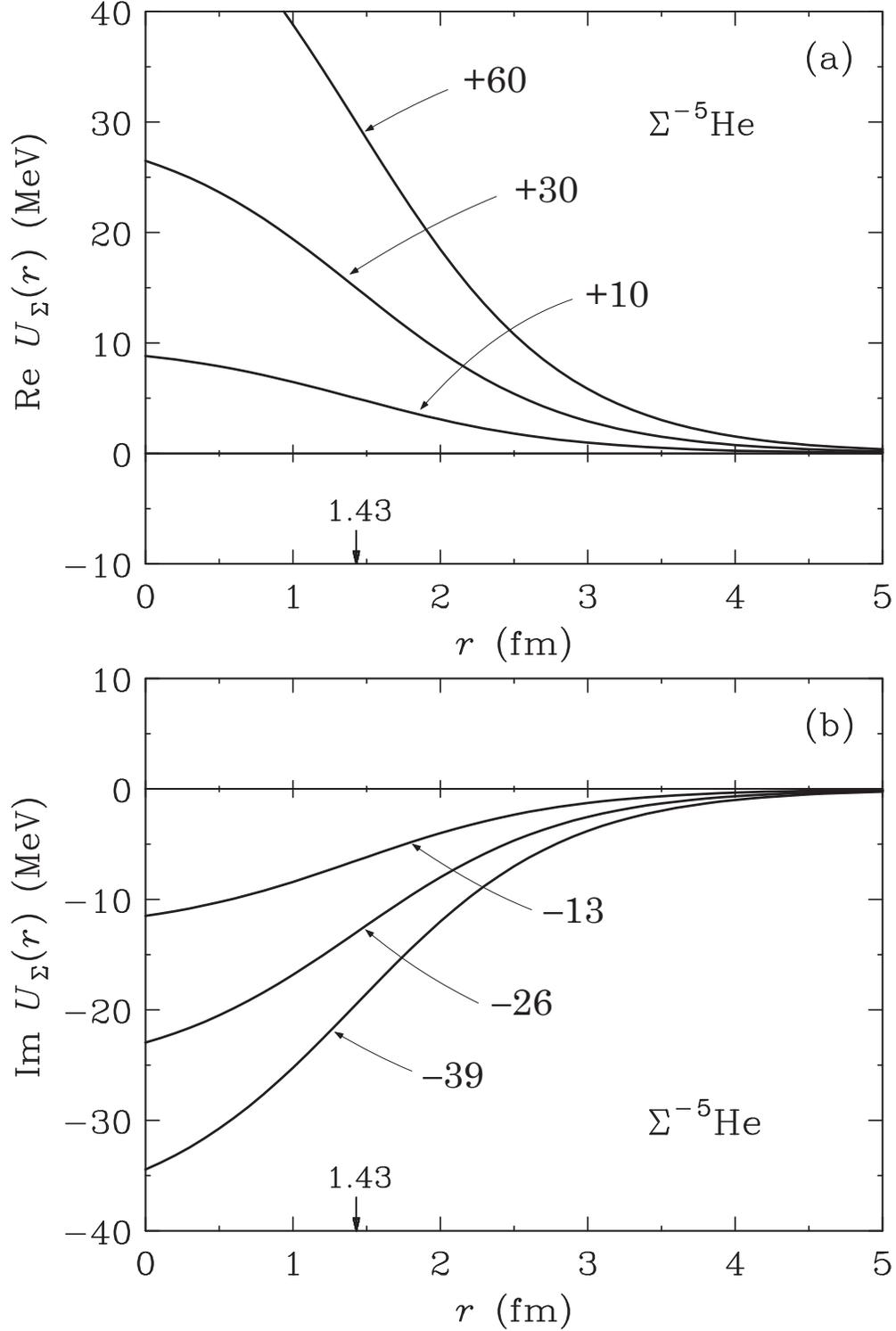}
  \end{center}
  \caption{\label{fig:5}
  (a) Real and (b) imaginary parts of the $\Sigma$-nucleus potentials 
  $U_{\Sigma}$ for $\Sigma^-$-$^{5}$He. 
  Solid curves denote the WS potential which has the strength of 
  $V_{\Sigma}=$ $+60$, $+30$, and $+10$ MeV and $W_{\Sigma}=$ $-$13, $-26$, and $-$39 MeV  
  with $R=$ $r_0A^{1/3}=$ 1.43 fm where $r_0=$ 0.835 fm and $a$= 0.706 fm. 
  }
\end{figure}

Consequently, we attempt to determine the values of ($V_\Sigma$,~$W_\Sigma$) 
phenomenologically by fitting to 
the shape and magnitude of the continuum spectra in $\Lambda$ and 
$\Sigma$ regions from the data of the J-PARC E10 experiment. 
In Fig.~\ref{fig:5}, we show the real and imaginary parts of the 
$\Sigma$-nucleus potentials for $\Sigma^-$-$^5$He, 
choosing the best-fit strengths of $V_\Sigma$= $+$30 MeV 
and $W_\Sigma$ = $-$26 MeV to fully explain the data of 
the $^{6}$Li($\pi^-$,$K^+$) data, as we will discuss 
them in the following sections.

\Figuretable{FIG. 5}

\section{Results}
\label{results}

Now we calculate the inclusive $K^+$ spectra for $\Sigma^-$-$^5$He 
($^6_{\Sigma}$H; $T=3/2$) using the Green's function method \cite{Morimatsu94}
in order to be compared with the data of the ${^6{\rm Li}}$($\pi^-$,~$K^+$)
reaction at 
the incident $\pi^-$ momentum of $p_{\pi^-}=$ 1.2 GeV/$c$ and 
the $K^+$ forward-direction angles of $\theta_{\rm lab}=$ 2$^\circ$--14$^\circ$.
The average cross section $\bar{\sigma}_{2^\circ\mbox{-}14^\circ}$ 
is given by 
\begin{equation}
\bar{\sigma}_{2^\circ\mbox{-}14^\circ} \equiv
\int_{\theta_{\rm lab}= 2^\circ}^{\theta_{\rm lab}= 14^\circ} \!\!
\left({d^2\sigma \over dE d\Omega} \right) d\Omega \bigg/
\int_{\theta_{\rm lab}= 2^\circ}^{\theta_{\rm lab}= 14^\circ} \!
d\Omega
\label{eqn:e21}
\end{equation}
in the laboratory frame.
To make a fit to the spectral shape of the data, 
we will introduce a renormalization factor of $f_s$ 
into the absolute value of the calculated spectrum 
because the amplitude of $\overline{f}_{\pi^-p \to K^+\Sigma^-}$ 
would have some ambiguities.

\Figuretable{FIG. 6}

\Figuretable{TABLE I}

\Figuretable{FIG. 7}

\subsection{Average cross section}

\subsubsection{$\chi^2$ fitting}

We examine the dependence of the spectral shape 
on two important strength parameters of ($V_\Sigma$,~$W_\Sigma$) 
in the WS potential with $R=r_0A_{\rm core}^{1/3}=$ 1.428 fm and 
$a=$ 0.706 fm, comparing the calculated spectrum for 
$\bar{\sigma}_{2^\circ\mbox{-}14^\circ}$ 
with the data of the $^{6}$Li($\pi^-$,~$K^+$) reaction at 1.2 GeV/$c$ 
from the J-PARC E10 experiment \cite{Honda16}.
We obtain the values of $\chi^2$ for fits to the data points 
of $N=$ 66 in the missing mass $M_{\rm x}=$ 5790--5920 MeV/$c^2$, 
varying the strengths of ($V_\Sigma$,~$W_\Sigma$) and $f_s$;
we estimate the average cross section in Eq.~(\ref{eqn:e21}), 
calculating the spectra for $\theta_{\rm lab}=$ 2$^\circ$--14$^\circ$
in parameter region of $V_\Sigma=$ ($-$20)--(+80) MeV 
and $-W_\Sigma=$ $0$--60 MeV by a 5 MeV energy step. 
The 2 MeV energy step is taken in the estimation near 
the $\chi^2_{\rm min}$ point.

\begin{figure}[tbh]
  \begin{center}
  \includegraphics[width=1.0\linewidth]{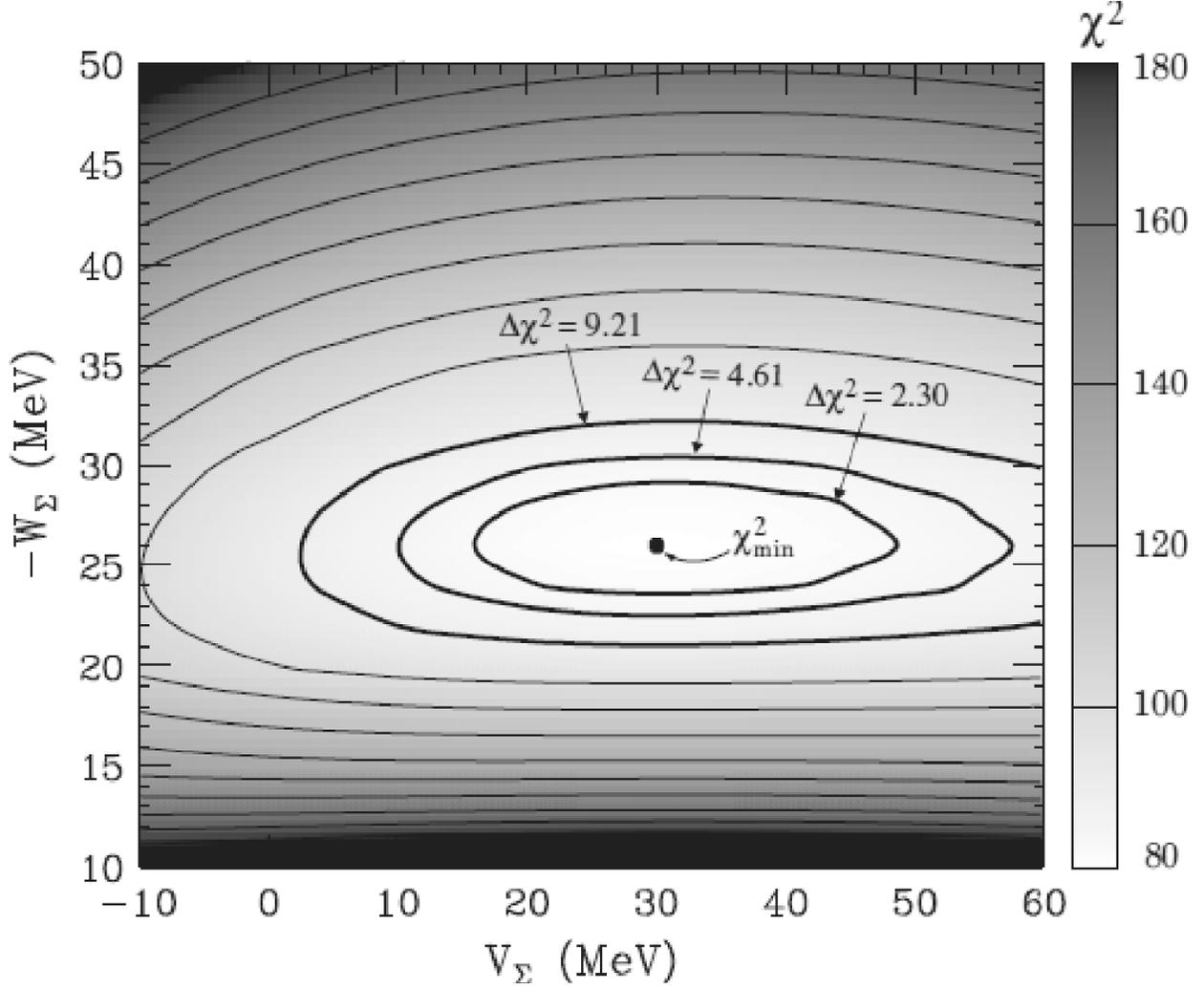}
  \end{center}
  \caption{\label{fig:6}
  Contour plots of the $\chi^2$-value distribution 
  in the \{$V_\Sigma$, $-W_\Sigma$\} plane from fitting to the average cross section 
  of $\bar{\sigma}_{2^\circ\mbox{-}14^\circ}$ in the $^6{\rm Li}$($\pi^-$,~$K^+$) 
  reaction at $p_{\pi^-}=$ 1.2 GeV/$c$. 
  A solid circle denotes the minimum position of $\chi^2_{\rm min}=$ 84.5 at 
 ($V_\Sigma$,~$-W_\Sigma$)= (30 MeV,~26 MeV) with $f_s=$ 1.23. 
  Thick curves indicate $\Delta \chi^2=$ 2.30, 4.61, and 9.21 which correspond  
  to 68\%, 90\%, and 99\% confidence levels for 2 parameters, respectively.
  }
\end{figure}

\begin{figure}[tbh]
\begin{center}
  \includegraphics[width=1.0\linewidth]{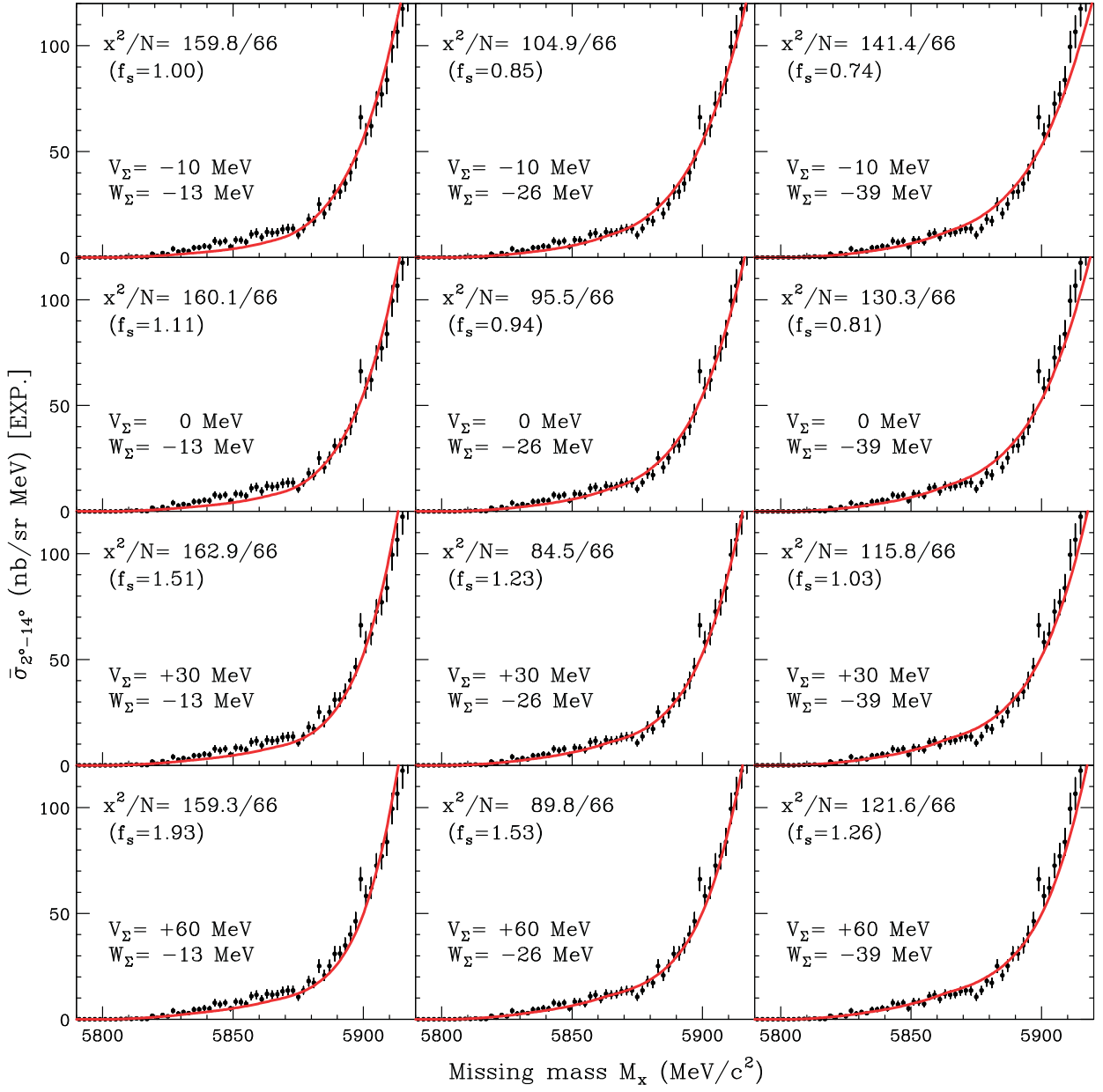}
\end{center}
\caption{\label{fig:7}
  Comparison of the calculated spectra for $\bar{\sigma}_{2^\circ\mbox{-}14^\circ}$ 
  with the data of the $^{6}$Li($\pi^-$,$K^+$) reaction 
  at $p_{\pi^-}$= 1.2 GeV/$c$~\cite{Honda16}.
  Solid curves denote the spectrum for the WS potential with ($V_\Sigma$, $W_\Sigma$)
  listed in Table \ref{tab:table1}, 
  together with the value of $\chi^2/N$ and the renormalization factor $f_s$. 
  The spectra are folded with a detector resolution of 3 MeV FWHM.
}
\end{figure}

Figure~\ref{fig:6} displays the contour plots of 
$\chi^2$-value distribution for $\bar{\sigma}_{2^\circ\mbox{-}14^\circ}$.
The minimum value of $\chi^2$ is found to be $\chi^2_{\rm min}=$ 84.5 
at $V_\Sigma =$ $+$30 MeV, $W_\Sigma =$ $-$26 MeV, and $f_s=$ 1.23, 
leading to elliptic regions of $\Delta \chi^2 =$ 2.30, 4.61, 
and 9.21 which correspond to 68\%, 90\%, and 99\% confidence levels 
for 2 parameters, respectively, where $\Delta \chi^2 \equiv \chi^2 - \chi^2_{\rm min}$.
Therefore, Figure~\ref{fig:6} clearly shows that a repulsive potential 
for $\Sigma^-$-$^5$He is needed to reproduce the data. 
In Table~\ref{tab:table1}, we show the reduced $\chi^2$ values 
of $\chi^2/N$ 
in calculations when $V_\Sigma$= $-$10, 0, $+$30, and $+$60 MeV 
and $W_\Sigma$ = $-$13, $-$26, and $-$39 MeV, comparing 
the calculated spectra with the data, as shown in Fig.~\ref{fig:7}. 
We find that the value of $\chi^2/N$ is fairly changed by $W_\Sigma$ 
and it is also dependent on $V_\Sigma$ 
although the visible difference of the fitting in Fig.~\ref{fig:7} 
does not seem to be so clear.
This analysis indicates that the shape and magnitude of the calculated spectrum 
are sensitive to ($V_\Sigma$,~$W_\Sigma$); 
the calculated spectrum for ($V_\Sigma$,~$W_\Sigma$) $\simeq$ ($+$30 MeV,~$-$26 MeV) 
is in good agreement with that of the data 
because it gives the minimum value of $\chi^2/N =$ 84.5/66= 1.28.

\begin{figure}[tbh]
\begin{center}
  \includegraphics[width=1.0\linewidth]{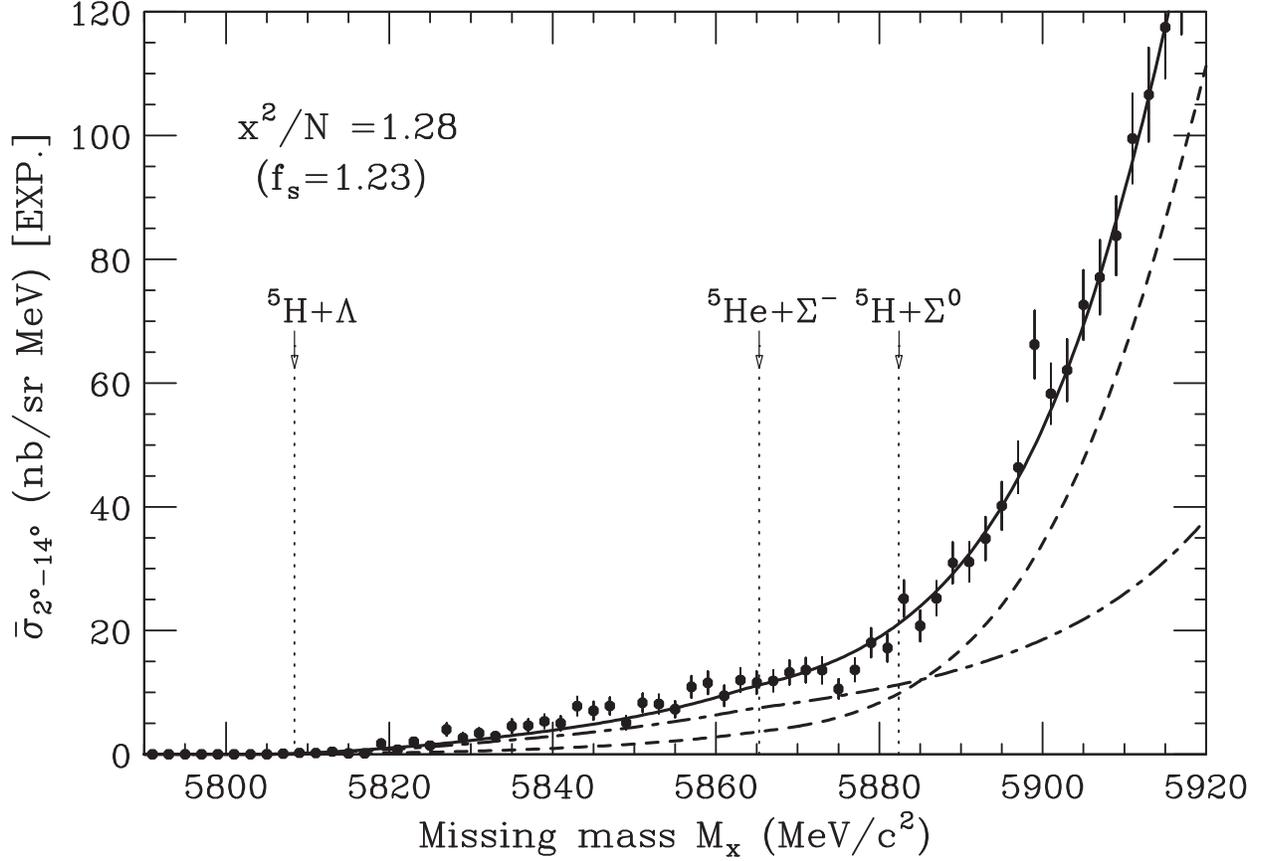}
\end{center}
\caption{\label{fig:8}
  Calculated spectrum for $\bar{\sigma}_{2^\circ\mbox{-}14^\circ}$ 
  in the WS potential with 
  $V_\Sigma=$ $+$30 MeV, $W_\Sigma=$ $-$26 MeV, $r_0=$ 0.835 fm, and 
  $a=$ 0.706 fm, together with the data of the $^{6}$Li($\pi^-$,$K^+$) 
  reaction at $p_{\pi^-}$= 1.2 GeV/$c$~\cite{Honda16}. 
  Solid, dot-dashed, and dashed curves denote total, $s$-hole, and $p$-hole 
  contributions, respectively. 
  The spectra are folded with a detector resolution of 3 MeV FWHM. 
}
\end{figure}

In Fig.~\ref{fig:8}, we show the absolute values of the calculated 
spectrum for $\bar{\sigma}_{2^\circ\mbox{-}14^\circ}$ 
with ($V_\Sigma$,~$W_\Sigma$)= (+30 MeV,~$-$26 MeV)
in comparison with the data for $M_{\rm x}=$ 5790--5920 MeV/$c^2$. 
We find the contribution of $p$-hole 
configurations is larger than that of $s$-hole configurations 
in the $\Sigma$ continuum region; 
the latter configurations dominate in the $\Lambda$ continuum region 
below the $^5$He+$\Sigma^-$ threshold, 
where the production strength mainly arises from a term of 
$G_{\Sigma}^\dagger({\rm Im}U_\Sigma)G_{\Sigma}$ 
describing the $\Sigma^-p \to \Lambda n$ processes in $^{5}$He
together with the core-nucleus breakup \cite{Harada98}. 
The optimal Fermi averaging for the $\pi^-p \to K^+ \Sigma^-$ reaction also 
indicates a good description of the energy in the 
$\Lambda$ and $\Sigma^-$ QF spectra in the nuclear ($\pi^-$,~$K^+$) reaction.

\Figuretable{FIG. 8}

\Figuretable{FIG. 9}

\subsubsection{Repulsion and absorption}

\begin{figure}[tbh]
\begin{center}
  \includegraphics[width=0.8\linewidth]{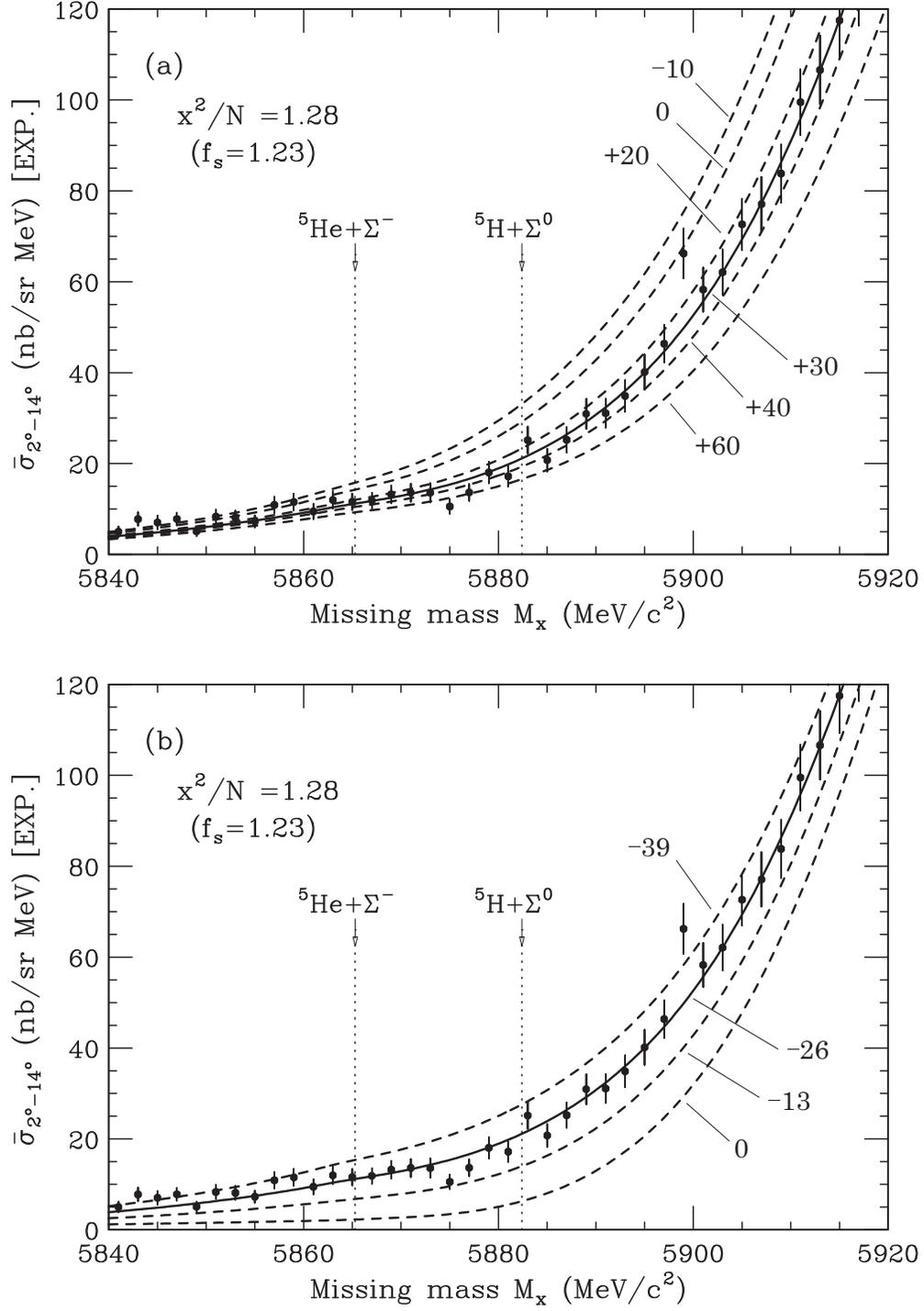}
\end{center}
\caption{\label{fig:9}
  Shapes and magnitudes of the calculated spectra for 
  $\bar{\sigma}_{2^\circ\mbox{-}14^\circ}$ in the $^{6}$Li($\pi^-$,$K^+$) reaction 
  at $p_{\pi^-}$= 1.2 GeV/$c$, depending on (a) the strengths of $V_\Sigma$ 
  when $W_\Sigma=$ $-$26 MeV is chosen  
  and (b) the strengths of $W_\Sigma$ when $V_\Sigma=$ $+$30 MeV is chosen.
  Solid curves denote the spectrum for the WS potential with  
  ($V_\Sigma$, $W_\Sigma$)= ($+$30 MeV, $-$26 MeV) as a guide.  
  The spectra are folded with a detector resolution of 3 MeV FWHM.
}
\end{figure}

\begin{figure}[tb]
\begin{center}
  \includegraphics[width=1.0\linewidth]{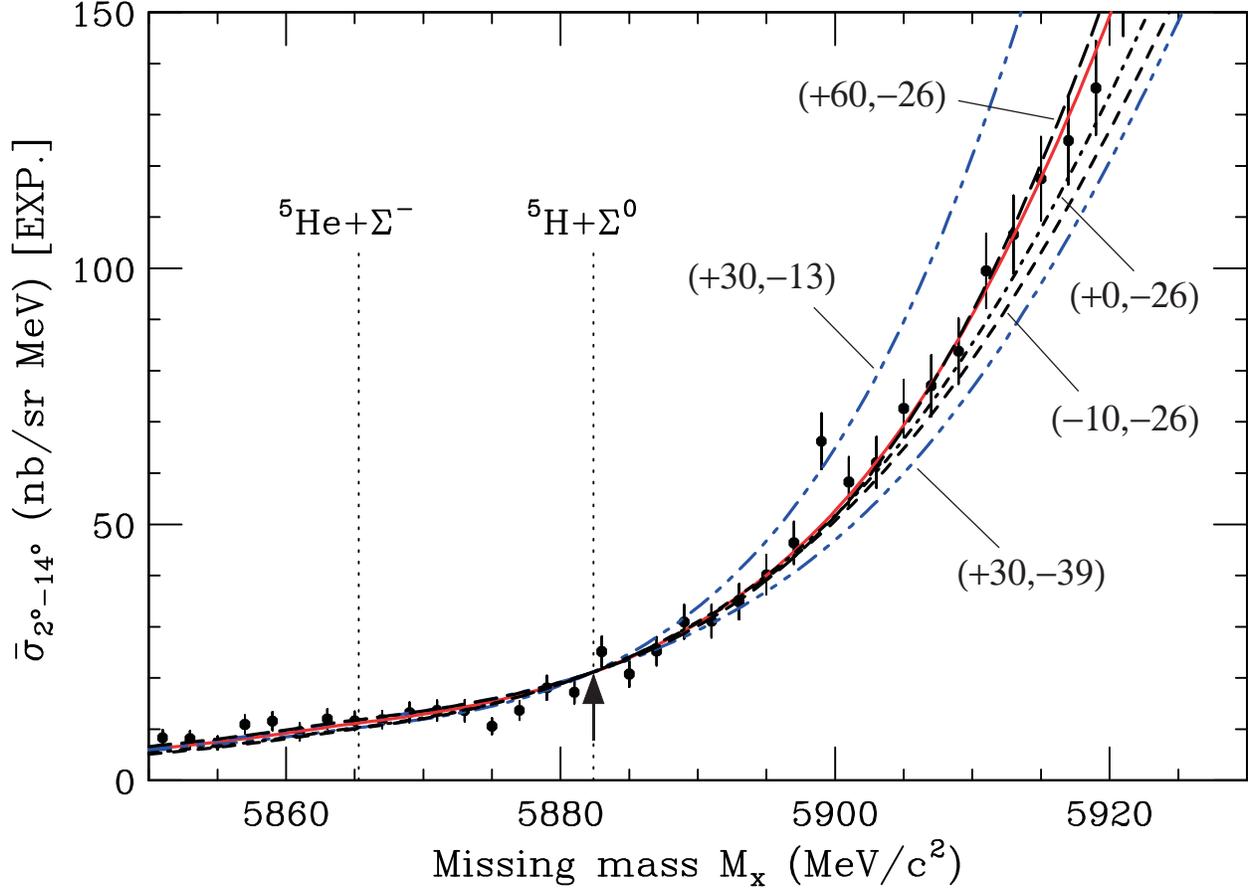}
\end{center}
\caption{\label{fig:10}
  Comparison among the shapes of the calculated spectra for 
  $\bar{\sigma}_{2^\circ\mbox{-}14^\circ}$
  in the $^{6}$Li($\pi^-$,$K^+$) reaction at $p_{\pi^-}$= 1.2 GeV/$c$, 
  together with the data~\cite{Honda16}.
  The solid curve denotes the spectrum with ($V_\Sigma$, $W_\Sigma$)
  =($+$30 MeV, $-$26 MeV) and $f_s=$ 1.23 as a guide.
  All the spectra for ($V_\Sigma$, $W_\Sigma$) are renormalized 
  at the point of the arrow
  corresponding to the $\Sigma^0$ threshold ($M_x=$ 5882.4 MeV/$c^2$) 
  to be compared among the shapes of them.
  The spectra are folded with a detector resolution of 3 MeV FWHM. 
}
\end{figure}

To see effects of the repulsion and absorption 
in the $\Sigma^-$-$^5$He potential, 
we discuss the shapes and magnitudes of the calculated spectra
depending on the strengths of $V_\Sigma$ and $W_\Sigma$. 
Figure~\ref{fig:9} shows the absolute values of the calculated 
spectra for $\bar{\sigma}_{2^\circ\mbox{-}14^\circ}$, 
using various strengths of $V_\Sigma$ and $W_\Sigma$ around 
the $\Sigma$ threshold. 
As seen in Fig.~\ref{fig:9}(a), the magnitude of the spectra 
in the $\Sigma^-$ region decreases and its slope becomes larger 
as the repulsion increases.
When the absorption increases, the yields in the $\Lambda$ 
region grow up, as shown in Fig.~\ref{fig:9}(b); 
the shape of the spectra trends to become flat, as a function of $M_{\rm x}$.
To clearly see a change of the shape of the spectrum on the parameter set, 
we display the calculated spectra for ($V_\Sigma$,~$W_\Sigma$), renormalizing
them for fits to the spectrum with 
($V_\Sigma$,~$W_\Sigma$)= (+30 MeV,~$-$26 MeV) at the $\Sigma^0$ threshold 
($M_x=$ 5882.4 MeV/$c^2$), as shown in Fig.~\ref{fig:10}.  
We find that the slope of the spectrum in the $\Sigma^-$ region is fairly enlarged, 
as a function of $M_x$, when the repulsion increases and the absorption decreases.
We recognize that the shape of the spectrum is significantly changed by 
the repulsion and absorption in the WS potential.  
Therefore, we confirm that the $\Sigma^-$-$^5$He potential
has a repulsion in the real part and a sizable absorption
in the imaginary part; the strengths denote 
\begin{eqnarray}
&V_\Sigma =& +30 \pm 10 \ {\rm MeV},  \nonumber \\
&W_\Sigma =& -26 \pm 2 \ {\rm MeV},
\label{eqn:e22}
\end{eqnarray}
in the WS potential with $r_0=$ 0.835 fm and $a=$ 0.706 fm, 
as seen in Fig.~\ref{fig:6}.
This potential provides the ability to explain 
the $^6$Li($\pi^-$,~$K^+$) data at the J-PARC E10 experiment,
although the radial shape of the WS potentials containing 
pure repulsion never explain the $\Sigma^-$ atomic 
data \cite{Harada05,Harada06}.

\Figuretable{FIG. 10}

\Figuretable{TABLE II}

\subsection{Angular distributions}

\begin{figure*}[tbh]
\begin{center}
  \includegraphics[width=1.0\linewidth]{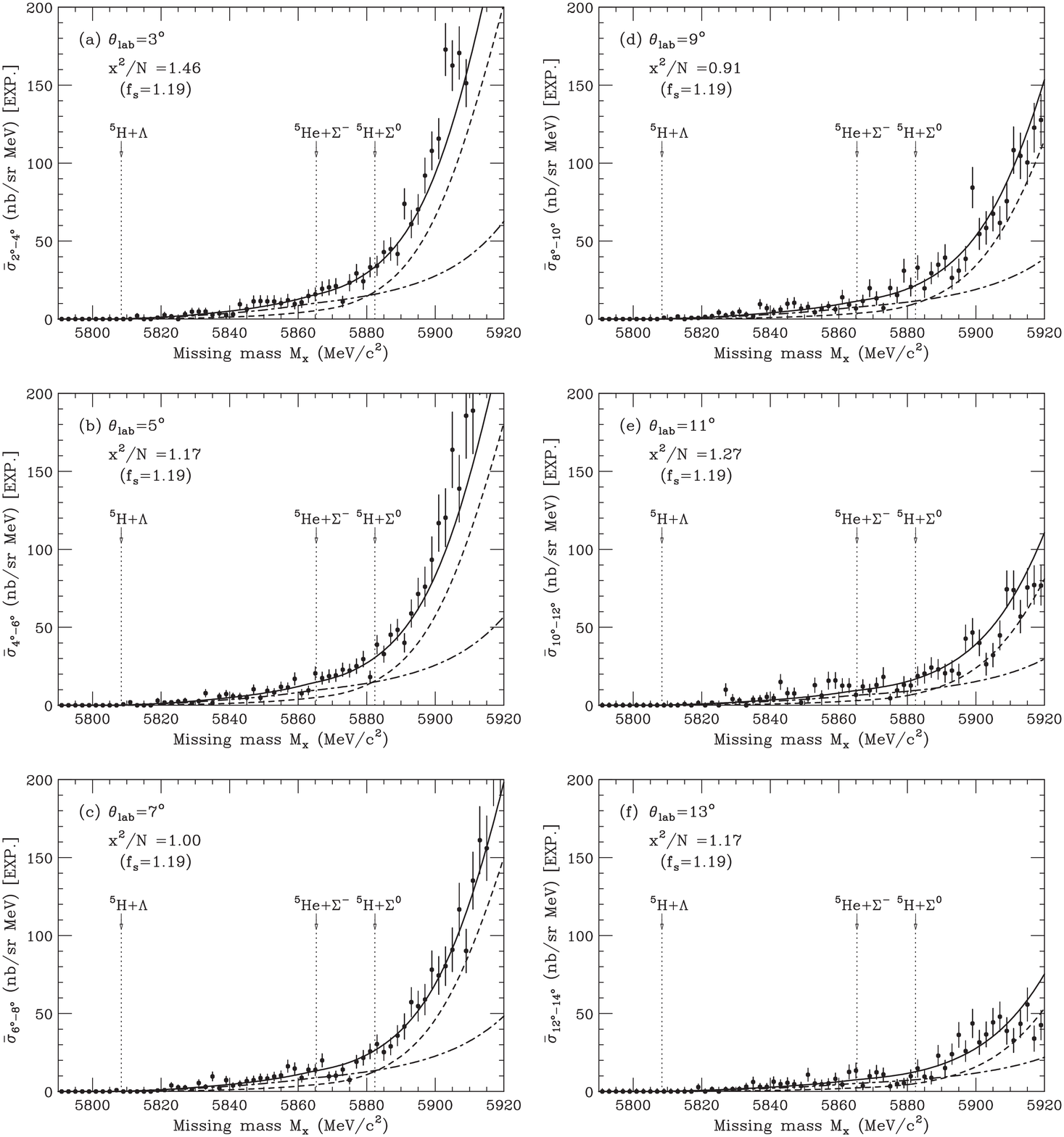}
\end{center}
\caption{\label{fig:11}
  Calculated spectra of the $^{6}$Li($\pi^-$,$K^+$) reaction 
  at $p_{\pi^-}$= 1.2 GeV/$c$ and $K^+$ forward-direction angles of 
  (a) $\theta_{\rm lab}=$ 3$^\circ$, (b) 5$^\circ$, (c) 7$^\circ$, (d) 9$^\circ$, 
  (e) 11$^\circ$, and (f) 13$^\circ$, as a function of the missing mass $M_{\rm x}$.
  The strengths of ($V_\Sigma$,~$W_\Sigma$)=($+$30 MeV, $-$26 MeV) are used 
  in the WS potential with $r_0=$ 0.835 fm and $a=$ 0.706 fm. 
  Solid, dot-dashed, and dashed curves denote total, $s$-hole, and $p$-hole contributions, 
  respectively, where the calculated spectra are normalized by a common factor $f_s=$ 1.19.
  The spectra are folded with a detector resolution of 3 MeV FWHM. 
  The data are taken from Ref.~\cite{Honda16}.
}
\end{figure*}

Figure~\ref{fig:11} displays the angular distributions for the calculated 
spectra with ($V_\Sigma$,~$W_\Sigma$)$=$ (+30 MeV,~$-$26 MeV)
at $\theta_{\rm lab}=$ 3$^\circ$, 5$^\circ$, 7$^\circ$, 
9$^\circ$, 11$^\circ$ and 13$^\circ$ for the missing mass 
$M_{\rm x}=$ 5790--5920 MeV/$c^2$ at 1.2 GeV/$c$. 
In Table~\ref{tab:table2}, we show the results of the values of $\chi^2$
for fits to the data of the angular distributions of 
$\theta_{\rm lab}=$ 2$^\circ$--14$^\circ$ \cite{Honda16}.
We confirm that the shapes and magnitudes of the calculated 
spectra with $V_\Sigma=$ $+30 \pm 10$ MeV and 
$W_\Sigma =$ $-26 \pm 2$ MeV are almost consistent with those of the data,
i.e., $\chi^2_{\rm tot}/N_{\rm tot}=$ 459.6/396 = 1.16 
where $N_{\rm tot}=$ $66 \times 6 =$ 396, 
and a renormalization factor $f_s =$ 1.19 
depending on the absolute values of $\overline{f}_{\pi^-p \to K^+\Sigma^-}$.
However, we find that it is difficult to determine the parameters 
of ($V_\Sigma$,~$W_\Sigma$) by only fits to each datum of the 
angular distribution at $\theta_{\rm lab} =$ 9$^\circ$, 11$^\circ$,  
and 13$^\circ$ because the $\chi^2$ values are slightly sensitive to 
the parameters within the experimental errors;
the data of the forward angles of $\theta_{\rm lab} \le$ 8$^\circ$ 
are important to determine the parameters of the potential.

\Figuretable{FIG. 11}

\section{Discussion}
\label{discussion}

\subsection{Potential strengths}

To see the validity of the $\Sigma^-$-$^5$He potential given in Eq.~(\ref{eqn:e22}), 
we consider the dependence of $A_{\rm core}$ on potential parameters 
in the $\Sigma$-nucleus potential.
The folding model shows that the strength in Eq.~(\ref{eqn:e13}) 
is written as 
\begin{eqnarray}
V_\Sigma+iW_\Sigma &=& 
[\bar{v}_{\Sigma N}+i\bar{w}_{\Sigma N}g(E_\Lambda)]  \nonumber\\
&& \times (\sqrt{\pi}a_{\Sigma N})^3{F(0)},
\label{eqn:e24}
\end{eqnarray}
where $F(0)$ is the form factor at the nuclear center in Eq.~(\ref{eqn:e18}).
When we use the strengths of ($V_\Sigma$, $W_\Sigma$) in Eq.~(\ref{eqn:e22}), 
$a_{\Sigma N}=$ 1.2 fm, and $F(0)=$ 0.103 fm$^{-3}$ for $A_{\rm core}=$ 5, we find
\begin{eqnarray}
&\bar{v}_{\Sigma N}&= +30 \pm 10 \ {\rm MeV},\nonumber \\
&\bar{w}_{\Sigma N}&= -26 \pm 2 \ {\rm MeV}. 
\label{eqn:e25}
\end{eqnarray}
Using Eq.~(\ref{eqn:e25}) in the folding model, we obtain the $\Sigma^-$-{$^{27}$Al} potential
in which the strengths of 
\begin{eqnarray}
&V_\Sigma & = +37 \pm 12 \ {\rm MeV}, \nonumber \\
&W_\Sigma & = -32 \pm 3 \ {\rm MeV},
\label{eqn:e26}
\end{eqnarray}
are determined with the WS form 
having $R=1.1A^{1/3}_{\rm core}$, $a=$ 0.67 fm, and $F(0)=$ 0.127 fm$^{-3}$ 
for $A_{\rm core}=$ 27.
We find that 
the strength of $V_\Sigma$ in Eq.~(\ref{eqn:e26}) is slightly larger than 
that of $V_\Sigma=$ (+20)--(+30) MeV for $\Sigma^-$-$^{27}$Al \cite{Harada05}, 
whereas 
the strength of $W_\Sigma$ in Eq.~(\ref{eqn:e26}) is as large as 
that of $W_\Sigma =$ ($-$20)--($-$40) MeV for $\Sigma^-$-$^{27}$Al \cite{Harada05}. 
We believe that 
the repulsive and absorptive components of 
the $\Sigma^-$-$^5$He potential are consistent with those of 
the $\Sigma^-$-$^{27}$Al potential quantitatively. 
If we extend the folding model to the nuclear matter (n.m.), 
we obtain the s.p.~potential as
\begin{eqnarray}
U_\Sigma{({\rm n.m.})}
&=& \int {v}_{\Sigma N}({\bm r}-{\bm r'}) \nonumber\\
&& \times 
\sum_{|k| < k_F,s,t}\left|{1 \over (2 \pi)^{3 \over 2}}{\rm e}^{i{\bm k}{\bm r'}}
\eta_s\eta_t \right|^2d{\bm r'}  \nonumber\\
&=& (\bar{v}_{\Sigma N}+i\bar{w}_{\Sigma N})(\sqrt{\pi}a_{\Sigma N})^3{2 k^3_F \over 3 \pi^2}, 
\label{eqn:e27}
\end{eqnarray}
where $\eta_s$ ($\eta_t$) is a spin (isospin) function for a nucleon, and $k_F$ is 
the Fermi momentum of 1.36 fm$^{-1}$. 
Using Eq.~(\ref{eqn:e25}), we find 
\begin{eqnarray}
&{\rm Re}\ U_\Sigma{({\rm n.m.})} & = +48 \pm 16 \ {\rm MeV}, \nonumber \\
&{\rm Im}\ U_\Sigma{({\rm n.m.})} & =  -42 \pm 3 \ {\rm MeV},
\label{eqn:e28}
\end{eqnarray}
which should be regarded as the depths of the $\Sigma^-$ s.p.~potential 
in nuclear matter of $(N-Z)/(N+Z)$= 0.2 at the normal density.
Therefore, we show that the $\Sigma^-$-$^5$He potential has 
the repulsion in the real part with the sizable imaginary part 
involving uncertainty within the experimental errors, 
consistent with the results in the previous works \cite{Harada05,Harada06}.

\subsection{Volume integral}

To evaluate the repulsion and absorption in the potentials, we show a volume 
integral per nucleon of the potential \cite{Harada05}, which is defined by
\begin{eqnarray}
J_R+ i J_I &=& {1 \over A_{\rm core}}\int U_\Sigma(r) d{\bm r}.
\label{eqn:e23}
\end{eqnarray}
For the $\Sigma^-$-$^5$He potential, we obtain 
\begin{eqnarray}
(J_R,~J_I)\simeq (257 \ \mbox{MeV$\,$fm$^3$},\ -222 \ \mbox{MeV$\,$fm$^3$}), 
\label{eqn:e23a}
\end{eqnarray}
using ($V_\Sigma$,~$W_\Sigma$)= (+30 MeV,~$-$26 MeV) 
with $r_0=$ 0.835 fm and $a=$ 0.706 fm,
in comparison with $(J_R,~J_I)\simeq$ (236 MeV$\,$fm$^3$,~$-$314 MeV$\,$fm$^3$)
for the $\Sigma^-$-$^{27}$Al potential with ($V_\Sigma$,~$W_\Sigma$)= (+30 MeV,~$-$40 MeV).
We find that the value of $J_R$ for $\Sigma^-$-$^5$He is almost similar to that 
for $\Sigma^-$-$^{27}$Al, rather than the value of $J_I$.
The value of $J_I$ for $\Sigma^-$-$^5$He is as large as that 
for $\Sigma^-$-$^{27}$Al by a factor of 0.7. 
This seems to originate from 
the nuclear structure of the $\alpha + n$ cluster or the unsaturation density 
for $^5$He 
because the volume integral is fairy affected by a $k_F$ dependence of 
the effective $\Sigma N$ interaction in the nucleus.
To clearly understand the repulsion and absorption in the potential, 
we need more theoretical investigations based on microscopic description.

\Figuretable{TABLE III}

\subsection{Size and potential range}

The size and shape of the folding-model potential depends on 
the range of the two-body force.  
Here we discuss the parameters of ($r_0$,~$a$) in the $\Sigma^-$-$^5$He potentials 
adjusted to give a best least-squares fit to the radial shape of the form factors 
obtained by folding several ranges of the $\Sigma N$ force 
into the matter MHO density distribution, 
following the procedure in Sect.~\ref{Sigmapot}.
Considering $a_{\Sigma N}=$ 0.8, 1.2, and 1.6 fm as a gaussian range in Eq.~(\ref{eqn:e17}), 
we obtain the WS potentials with the adjusted parameters of ($r_0$,~$a$), 
and show the corresponding values of $\chi^2$ for the best fit to the data 
of $\bar{\sigma}_{2^\circ\mbox{-}14^\circ}$ in Table~\ref{tab:table3}; 
the value of $\chi^2_{\rm min}=$ 84.5 when $a_{\Sigma N}=$ 1.2 fm
is minimum in comparison with $\chi^2_{\rm min}=$ 84.7 (86.1) when $a_{\Sigma N}=$ 0.8 (1.6) fm.
We repeat that for $a_{\Sigma N}=$ 1.2 fm 
the strengths of ($V_\Sigma$,~$W_\Sigma$)= ($+$30 MeV,~$-$26 MeV) are 
favored for fits to the data, using the WS potential 
with $r_0=$ 0.835 fm and $a=$ 0.706 fm, 
as already shown in Table~\ref{tab:table1}.
We stress that the potential parameters for $\Sigma^-$-$^5$He should be carefully adopted
due to the unsaturation properties of the light nuclear core,  
as discussed in Sect.~\ref{Folding}.

On the other hand, the calculated spectra are rather insensitive to 
the potential with the parameters of ($r_0$, $a$) 
that give the similar values of $J_R+iJ_I$ 
with the best-fit ($V_{\Sigma}$,~$W_{\Sigma}$). 
In a previous work for $\Sigma^-$-$^5$He \cite{Harada17}, 
we used the WS potential with $r_0=$ 1.1 fm and $a=$ 0.67 fm, 
of which parameters were used in the analysis of the ($\pi^-$,~$K^+$) 
reactions on the heavier targets \cite{Harada17}.
We obtained $\chi^2_{\rm min}=$ 87.6 when the best-fit 
($V_\Sigma$,~$W_\Sigma$)= (+20 MeV,~$-$20 MeV). 
Because the value of $r_0=$ 1.1 fm seems to be too large for 
$A_{\rm core}=$ 5, 
the $\Sigma^-$-$^5$He potential should be improved by 
($V_\Sigma$,~$W_\Sigma$)= ($+$30 MeV,~$-$26 MeV) with 
$r_0=$ 0.835 fm and $a=$ 0.706 fm, as shown in Fig.~\ref{fig:5}. 
Nevertheless, one expects that the $\Sigma^-$ wave functions related 
to the contribution of $\Sigma\Lambda$ couplings in $^6_\Lambda$H 
are not modified 
because the previous potential with $r_0=$ 1.1 fm and $a=$ 0.67 fm 
gives 
the similar volume integral of Eq.~(\ref{eqn:e23a}), i.e., 
($J_R$,~$J_I$)= (253 MeV$\,$fm$^3$, $-$253 MeV$\,$fm$^3$)
for ($V_\Sigma$,~$W_\Sigma$)= (+20 MeV,~$-$20 MeV).

\subsection{Angular dependence of Fermi-averaged amplitudes}

In Table~\ref{tab:table2}, we showed the results of the values of $\chi^2$
for fits to the data of the angular distributions 
at $\theta_{\rm lab}=$ 3$^\circ$, 5$^\circ$, 7$^\circ$, 
9$^\circ$, 11$^\circ$, and 13$^\circ$.
We realized that $V_\Sigma=$ $+30 \pm 10$ MeV and 
$W_\Sigma =$ $-26 \pm 2$ MeV are favored to reproduce the data of 
the angular distributions, 
so that $\chi^2_{\rm tot}=$ 459.6 and a common factor $f_s =$ 1.19 are determined.
However, the angular distributions usually depend on the Fermi-averaged amplitudes of 
$\overline{f}_{\pi^-p \to K^+\Sigma^-}$ at $\theta_{\rm lab}$
as well as properties of the $\Sigma$-nucleus potentials.
Here we test the angular dependence of $\overline{f}_{\pi^-p \to K^+\Sigma^-}$
by introducing the renormalization factors of $f_{s,\theta}$ at each angle 
$\theta_{\rm lab}$, rather than a common factor of $f_{s}$. 
In Table~\ref{tab:table4}, we show the results of $\chi^2/N$ values 
for fits to the data at each $\theta_{\rm lab}$
when we take ($V_\Sigma$,~$W_\Sigma$)= ($+$30 MeV,~$-$26 MeV). 
We find $\chi^2_{\rm tot}/N_{\rm tot}=$ 388.5/396= 0.981, which is significantly improved 
by each $f_{s,\theta}$ in comparison with $\chi^2_{\rm tot}/N_{\rm tot}=$ 459.6/396= 1.16, 
as seen in Table~\ref{tab:table2};
the absolute values of $\overline{f}_{\pi^-p \to K^+\Sigma^-}$ 
at $\theta_{\rm lab}=$ 3$^\circ$ and 5$^\circ$ are enlarged by 13\% 
and 7\%, respectively, whereas those at $\theta_{\rm lab}=$ 
7$^\circ$, 9$^\circ$, 11$^\circ$, and 13$^\circ$ are reduced 
by 10\%, 6\%, 8\%, and 16\%, respectively. 
This improvement may suggest that the angular distributions 
of the optimal Fermi-averaged amplitudes of $\overline{f}_{\pi^-p \to K^+\Sigma^-}$ 
still have some ambiguities, 
as well as the simplicity of the s.p.~shell-model description 
for $^6$Li and $^5$He in our DWIA calculations.

\Figuretable{TABLE IV}

\subsection{Neutron-excess environment}

The neutron-rich nuclei give us new information on properties of the nuclear 
structure and two-body $NN$ force because of unusual behaviors 
of the excess neutrons such as neutron skin and neutron halo.
The sizes of the neutron-rich He and Li isotopes are also 
discussed experimentally and theoretically \cite{Tanihata13}.
The study of neutron-rich $\Sigma$ hypernuclei is one of the most 
promising subjects to examine the hypernuclear potentials 
in the neutron-excess environment. 
In this work, the DCX reaction ($\pi^-$,~$K^+$) on the $^6$Li($1^+$; $T=0$) target 
provides a population of 
the neutron-rich $\Sigma^-$-$^5$He and $\Lambda$-$^5$H hypernuclei with $T=$ 3/2, 
where effects of the potential strengths on the neutron-excess
environments of $(N-Z)/(N+Z)=$ 0.2--0.6 are expected to be enhanced. 

Our analyses for the $\Sigma^-$-$^5$He potential suggest 
the strength of ${\rm Re}\ U_\Sigma \simeq$ $+48 \pm 16$ MeV
extrapolated to neutron-excess matter at the normal density, 
which is larger than that of 
${\rm Re}\ U_\Sigma \simeq$ ($+20$)--($+30$) MeV 
obtained by usual $N \simeq Z$ nuclei. 
We show that the $\Sigma^-$-$^5$He potential becomes 
more repulsive in the real part because the repulsion of $\Sigma N$ $I=$ 3/2, 
$^3$S$_1$ increases in $\Sigma^-$+$^5$He with the neutron-excess 
environments of $(N-Z)/(N+Z)=$ 0.2, 
and that it has a sizable absorption in the imaginary part 
because a conversion transition to continuum nuclear breakup states in 
$\Lambda$+$^5$H$^*$ with $(N-Z)/(N+Z)=$ 0.6 would be enlarged.

However, it should be noticed that nuclear effects of the 
$\alpha + d$ cluster structure 
and the nuclear deformation in light nuclei, and the nuclear coupled channels 
in $\Lambda$-$^5{\rm H}$ continuum states \cite{Harada14}
are not taken into account in our calculations. 
More theoretical investigations based on microscopic description 
are needed to clarify the nature of the $\Sigma$-nucleus potential.

\section{Summary and Conclusion}
\label{summary}

We have studied phenomenologically the inclusive spectra of 
the $^6$Li($\pi^-$,~$K^+$) reaction at $p_{\pi^-}=$ 1.2 GeV/$c$ 
within the DWIA in order to clarify the property of the $\Sigma$-nucleus potential
for $\Sigma^-$-$^5$He. 
We have determined the strengths of ($V_\Sigma$,~$W_\Sigma$) in the 
WS potential with $r_0=$ 0.835 fm and $a=$ 0.706 fm by comparing the calculated 
continuum spectrum in $\Sigma$ and $\Lambda$ regions 
with the data of the J-PARC E10 experiment. 
We have also discussed effects of the size and potential range for $\Sigma^-$-$^5$He
in the neutron excess of $(N-Z)/(N+Z)=$ 0.2. 
The results are summarized as follows:
\begin{itemize}
\item[(i)]
The calculated spectra with DWIA can fully reproduce the data of 
$\bar{\sigma}_{2^\circ\mbox{-}14^\circ}$ and the angular distribution 
at $\theta_{\rm lab}=$ 2$^\circ$--14$^\circ$ in 
the $^6$Li($\pi^-$,~$K^+$) reaction at 1.2 GeV/$c$.
\item[(ii)]
The repulsive and absorptive components for the $\Sigma^-$-$^5$He potential 
indicate $V_\Sigma=$ $+30 \pm 10$ MeV and $W_\Sigma=$ $-26 \pm 4$ MeV 
in the WS potential so as to explain the data of the J-PARC E10 experiment.
\item[(iii)]
The optimal Fermi-averaged amplitudes of $\overline{f}_{\pi^-p \to K^+\Sigma^-}$ 
in our DWIA calculations
are essential to describe the energy and angular dependence of the data 
of the $^6$Li($\pi^-$,~$K^+$) reaction at 1.2 GeV/$c$. 
\end{itemize}

In conclusion, we show that 
the repulsive and absorptive components of the $\Sigma^-$-$^5$He 
potential provide the ability to explain 
the data of the $^6$Li($\pi^-$,~$K^+$) spectra at 1.2 GeV/$c$; 
the strengths of $V_\Sigma \simeq$ +30 MeV and $W_\Sigma \simeq$ $-$26 MeV 
are favored within the WS potential, consistent with analyses for heavier nuclei. 
We recognize that the calculated spectra via $\Sigma^-$ doorways 
can reproduce the experimental data of the $^{6}$Li($\pi^-$,~$K^+$) reaction 
at 1.2 GeV/$c$ in $\Sigma$ region as well as $\Lambda$ region \cite{Harada17}. 
The detailed analysis based on microscopic calculations 
is required for the analyses of the J-PARC E10 experiment.
This investigation is in progress. 
\\

\begin{acknowledgments}
The authors would like to thank Professor~A.~Sakaguchi, 
Professor~H.~Tamura, Dr.~M.~Ukai and Professor~T.~Fukuda 
for many valuable discussions.
This work was supported by JSPS KAKENHI Grants 
No.~JP24105008, No.~JP25400278, and No.~JP16K05363.
\end{acknowledgments}


\clearpage
\begin{table}[bth]
\caption{
\label{tab:table1}
The $\chi^2$-fitting for various strength parameters, $V_\Sigma$ and $W_\Sigma$, 
in the WS potential for $\Sigma^-$-$^{5}$He, where $r_0=$ 0.835 fm and $a=$ 0.706 fm.
The value of $\chi^2/N$ and the renormalization factor $f_s$ are obtained 
by comparing the calculated spectrum with the $N=$ 66 data points 
of the average cross sections of $\bar{\sigma}_{2^\circ\mbox{-}14^\circ}$ 
for the missing mass $M_{\rm x}$= 5790--5920 MeV/$c^2$.
The data were taken from Ref.~\cite{Honda16}.
}
\begin{ruledtabular}
\begin{tabular}{ccrc}
\noalign{\smallskip}
  $V_\Sigma$ & $W_\Sigma$ 
  &  \multicolumn{2}{c}{$\bar{\sigma}_{2^\circ\mbox{-}14^\circ}$}   \\
\noalign{\smallskip}
       \cline{3-4}  
\noalign{\smallskip}
  (MeV)    &  (MeV)  & $\chi^2/N\ \ $   & $f_s$    \\
\noalign{\smallskip}\hline\noalign{\smallskip}
   $-10$     &   $-13$  & 159.8/66  & 1.00  \\
   \ \ $0$   &   $-13$  & 160.1/66  & 1.11  \\
   $+30$     &   $-13$  & 162.9/66  & 1.51  \\
   $+60$     &   $-13$  & 159.3/66  & 1.93  \\
   $-10$     &   $-26$  & 104.9/66  & 0.85  \\
   \ \ $0$   &   $-26$  &  95.5/66  & 0.94  \\
   $+30$     &   $-26$  &  84.5/66  & 1.23  \\
   $+60$     &   $-26$  &  89.8/66  & 1.53  \\
   $-10$     &   $-39$  & 141.4/66  & 0.74  \\
   \ \ $0$   &   $-39$  & 130.3/66  & 0.81  \\
   $+30$     &   $-39$  & 115.8/66  & 1.03  \\
   $+60$     &   $-39$  & 121.6/66  & 1.26  \\
\end{tabular}
\end{ruledtabular}
\end{table}

\begin{table*}[bth]
\caption{
\label{tab:table2}
The $\chi^2$-fitting for various strength parameters, $V_\Sigma$ and $W_\Sigma$, 
in the WS potential for $\Sigma^-$-$^{5}$He, where $r_0=$ 0.835 fm and $a=$ 0.706 fm.
The value of $\chi^2$ and the common renormalization factor $f_s$ are obtained 
by comparing the calculated spectrum with the $N=$ 66 data points of 
the angular distributions at $\theta_{\rm lab}=$
3$^\circ$, 5$^\circ$, 7$^\circ$, 9$^\circ$, 11$^\circ$,
and 13$^\circ$ for the missing mass $M_{\rm x}$= 5790--5920 MeV/$c^2$.
The data were taken from Ref.~\cite{Honda16}.
}
\begin{ruledtabular}
\begin{tabular}{cccccccccc}
\noalign{\smallskip}
     $V_\Sigma$ & $W_\Sigma$  
     &  \multicolumn{6}{c}{$\chi^2$}  
     &  $\chi^2_{\rm tot}$\tablenotemark[1]
     &  $f_s$ \\
\noalign{\smallskip}
       \cline{3-8}  
\noalign{\smallskip}
     (MeV)        & (MeV)      & {$\ 2^\circ$-$\ 4^\circ$}
                               & {$\ 4^\circ$-$\ 6^\circ$}
                               & {$\ 6^\circ$-$\ 8^\circ$}  
                               & {$\ 8^\circ$-$10^\circ$}
                               & {$10^\circ$-$12^\circ$}
                               & {$12^\circ$-$14^\circ$}
                               &  
                               &  \\
\noalign{\smallskip}\hline\noalign{\smallskip}
 $-10$     &   $-13$ &  88.4     & 87.3   & 74.9   & 83.7   & 106.8  & 87.3  & 492.3  & 0.97 \\
 \ \ $0$   &   $-13$ &  78.0     & 83.2   & 70.5   & 85.9   & 107.8  & 88.2  & 496.8  & 1.08 \\
 $+30$     &   $-13$ &  67.2     & 77.4   & 65.9   & 88.6   & 105.8  & 88.4  & 618.5  & 1.47 \\
 $+60$     &   $-13$ &  70.3     & 76.6   & 64.9   & 85.1   & 99.3   & 84.8  & 478.6  & 1.89 \\
 $-10$     &   $-26$ &  136.2    & 92.6   & 82.1   & 61.4   & 85.5   & 75.4  & 469.7  & 0.82 \\
 \ \ $0$   &   $-26$ &  120.8    & 86.1   & 75.2   & 60.7   & 85.4   & 75.9  & 590.4  & 0.91 \\
 $+30$     &   $-26$ &  96.6     & 77.2   & 65.7   & 60.1   & 83.7   & 77.0  & 459.6  & 1.19 \\
 $+60$     &   $-26$ &  94.5     & 79.2   & 66.6   & 60.2   & 81.1   & 77.3  & 428.9  & 1.50 \\
 $-10$     &   $-39$ &  210.5    & 124.2  & 111.2  & 64.4   & 79.3   & 74.2  & 545.3  & 0.71 \\
 \ \ $0$   &   $-39$ &  194.8    & 118.1  & 104.7  & 62.6   & 78.9   & 74.6  & 448.4  & 0.78 \\
 $+30$     &   $-39$ &  167.0    & 109.2  & 94.4   & 60.6   & 77.7   & 76.2  & 427.6  & 0.99 \\
 $+60$     &   $-39$ &  160.6    & 110.6  & 94.3   & 61.6   & 77.0   & 78.0  & 542.5  & 1.23 \\
\end{tabular}
\end{ruledtabular}
\tablenotetext[1]{
$\chi^2_{\rm tot}= \sum \chi^2$ for all data points of $N_{\rm tot}= 66 \times 6 = 396$.}
\end{table*}

\begin{table*}[bth]
\caption{
\label{tab:table3}
Effects of a gaussian range $a_{\Sigma N}$ in the two-body $\Sigma N$ force
on the $\Sigma^-$-$^5$He potential.
The values of $\chi^2$ are obtained by fits to the data of the average cross section 
for $\bar{\sigma}_{2^\circ\mbox{-}14^\circ}$.
The data were taken from Ref.~\cite{Honda16}.
}
\begin{ruledtabular}
\begin{tabular}{cccccccccc}
\noalign{\smallskip}
 $a_{\Sigma N}$ 
 & $r_0$\tablenotemark[1]
 & $a$\tablenotemark[1]
 & $R$\tablenotemark[1]
 & $V_\Sigma$ 
 & $W_\Sigma$  
 & $J_R+iJ_I$\tablenotemark[2]
 & $\langle r^2 \rangle^{1/2}_U$\tablenotemark[3]
 & \multicolumn{2}{c}{$\bar{\sigma}_{2^\circ\mbox{-}14^\circ}$}\\
\noalign{\smallskip}
       \cline{9-10}
\noalign{\smallskip} 
 (fm)  
 & (fm)  
 & (fm)  
 & (fm)  
 & (MeV) 
 & (MeV) 
 & (MeV)  
 & (fm)
 & $\chi^2$ &  $f_s$  \\
\noalign{\smallskip}\hline\noalign{\smallskip}
 0.8  &  0.701 & 0.646 & 1.20  &  $+40$ & $-35$ & $+231+i(-202)$ &  6.56 &  84.7  & 1.235     \\
 1.2  &  0.835 & 0.706 & 1.43  &  $+30$ & $-26$ & $+257+i(-222)$ &  8.06 &  84.5  & 1.227     \\
 1.6  &  0.997 & 0.778 & 1.71  &  $+25$ & $-20$ & $+324+i(-259)$ &  10.1 &  86.1  & 1.168     \\
\end{tabular}
\end{ruledtabular}
\tablenotetext[1]{
Parameters of the WS form: $f(r)=[1+\exp{((r-R)/a)}]^{-1}$, 
where $R=r_0A^{1/3}_{\rm core}$.
}
\tablenotetext[2]{
$J_R+iJ_I= \int U(r)d{\bm r}/A_{\rm core}$.
}
\tablenotetext[3]{
$\langle r^2 \rangle^{1/2}_U= \int r^2U(r)d{\bm r}/\int U(r)d{\bm r}$.
}
\end{table*}

\begin{table*}[hbt]
\caption{
\label{tab:table4}
The $\chi^2$ values and the renormalization factors $f_{s,\theta}$ 
for each $\theta_{\rm lab}$ 
by comparing the calculated spectrum with the $N=$ 66 data points 
of the angular distributions for the missing mass $M_{\rm x}$= 5790--5920 MeV/$c^2$.
$V_\Sigma=$ +30 MeV and $W_\Sigma=$ $-$26 MeV are used in the WS potential.
The value in the bracket is a ratio of $f_{s,\theta}$ to $f_s=$ 1.19  
which is taken as a common renormalization factor.
}
\begin{ruledtabular}
\begin{tabular}{lccccccc}
\noalign{\smallskip}
  $\theta_{\rm lab}$   
     &  {$\ 2^\circ$-$\ 4^\circ$}  
     &  {$\ 4^\circ$-$\ 6^\circ$}  
     &  {$\ 6^\circ$-$\ 8^\circ$}  
     &  {$\ 8^\circ$-$10^\circ$}  
     &  {$10^\circ$-$12^\circ$}  
     &  {$12^\circ$-$14^\circ$} 
     &  Total\tablenotemark[1]  \\
\noalign{\smallskip}\hline\noalign{\smallskip}
$\chi^2/N$ & 68.6/66 & 70.6/66 & 65.3/66  & 56.1/66 & 65.3/66  & 62.7/66  & 388.5/396 \\
$f_{s,\theta}$   & 1.35 & 1.28 &	1.17 &	1.12 &	1.01 &	1.00  & \\
$(f_{s,\theta}/f_s)$   & (1.13) & (1.07) & (0.90) & (0.94) & (0.92) & (0.84)  & \\
\noalign{\smallskip}                                                                           
\end{tabular}
\end{ruledtabular}
\tablenotetext[1]{
$\chi^2_{\rm tot}/N_{\rm tot}$ where $N_{\rm tot}= 66 \times 6 = 396$.
}
\end{table*}

\clearpage

\end{document}